\documentclass{aa}
\newcommand{\ov}{\alpha_{\mathrm{ov}}}
\newcommand{\fov}{f_{\mathrm{ov}}}
\newcommand{\conv}{\alpha_{\mathrm{conv}}}
\newcommand{\dd}{\mathrm{d}}
\newcommand{\teff}{T_{\mathrm{eff}}}

\newcommand{\lsol}{L_{\odot}}
\newcommand{\nug}{\nu_{\mathrm{g}}}
\newcommand{\smass}{M_{\odot}}

\newcommand{\metal}{[\mathrm{Fe}/\mathrm{H}]}
\newcommand{\N}{Brunt-Väisälä }

\newcommand{\kepler}{\emph{Kepler}}
\newcommand\T{\rule{0pt}{2.6ex}}
\newcommand\B{\rule[-1.2ex]{0pt}{0pt}}
\usepackage[normalem]{ulem}
\usepackage[outdir=./figures]{epstopdf}

\usepackage{graphicx}
\usepackage{txfonts} 
\usepackage{natbib}

\begin{document}

\title{Probing core overshooting using subgiant asteroseismology: The case of KIC10273246}
\author{A. Noll\inst{1}
\and S. Deheuvels\inst{1}
\and J. Ballot\inst{1}}
\institute{IRAP, Universit\'e de Toulouse, CNRS, CNES, UPS, (Toulouse), France}

\offprints{A. Noll\\ \email{anoll@irap.omp.eu}}

\abstract{The size of convective cores remains uncertain, despite their substantial influence on stellar evolution, and thus on stellar ages. The seismic modeling of young subgiants can be used to obtain indirect constraints on the core structure during main sequence, thanks to the high probing potential of mixed modes.}
{We selected the young subgiant KIC10273246, observed by Kepler, based on its mixed-mode properties. We thoroughly modeled this star, with the aim of placing constraints on the size of its main-sequence convective core. A corollary goal of this study is to elaborate a modeling technique that is suitable for subgiants and can later be applied to a larger number of targets.}
{We first extracted the parameters of the oscillation modes of the star using the full Kepler data set. To overcome the challenges posed by the seismic modeling of subgiants, we propose a method that is specifically tailored to subgiants with mixed modes and uses nested optimization. We then applied this method to perform a detailed seismic modeling of KIC10273246.}
{We obtain models that show good statistical agreements with the observations, both seismic and non-seismic. We show that including core overshooting in the models significantly improves the quality of the seismic fit, optimal models being found for $\ov=0.15$. Higher amounts of core overshooting strongly worsen the agreement with the observations and are thus firmly ruled out. We also find that having access to two g-dominated mixed modes in young subgiants allows us to place stronger constraints on the gradient of molecular weight in the core and on the central density.}
{This study confirms the high potential of young subgiants with mixed modes to investigate the size of main-sequence convective cores. It paves the way for a more general study including the subgiants observed with Kepler, TESS, and eventually PLATO.}
\maketitle

\section{Introduction}

One of the most important current open questions in stellar physics is the extent of convective cores. Several physical processes are known to extend the convective core boundaries beyond the standard Schwarzschild limit. The most frequently quoted are overshooting of ascending blobs of fluids due to their inertia, rotational mixing or semi-convection. All these processes remain poorly described by theory, and the way they interact is understood even less. They are therefore generally modeled, in stellar evolution codes, as an extension of the mixed core over a distance $d_{\mathrm{ov}}$, which is often referred to as the distance of overshoot, even though other processes can contribute as well. In practice, this can either be achieved by simply extending the fully mixed central region, or by treating overshooting as a diffusive process, following a behavior found in numerical simulations \citep{freytag96}. In both cases, a free parameter controlling the extent of the additional mixing is required. Observations are therefore necessary to better constrain those processes.

Initial constraints have been obtained thanks to the study of the HR diagram of clusters (see e.g., \citealt{maeder89}, \citealt{vandenberg06}), and the modeling of eclipsing binaries \citep{claret16}. Most of those studies favor adding overshooting, to various extents. Typically, $d_{\mathrm{ov}}$ is around $0.2 \, H_p$, where $H_p$ is the pressure scale height. \cite{claret16} found that $\ov$, the ratio between $d_{\mathrm{ov}}$ and $H_p$, increases with mass for stars under 2 $\smass$ before reaching a plateau. However, this result is still debated (\citealt{constantino18}, \citealt{claret19}).

Over the last decade, asteroseismology allowed us to probe the structure of stellar cores.
Thanks to the data of CoRoT \citep{baglin06}, \textit{Kepler} \citep{borucki10} and now TESS \citep{ricker14} missions, we have been able to precisely measure the oscillation frequencies of numerous pulsators. 
The study of pressure (p) modes in low-mass main sequence (MS) stars, showed the need for core overshooting to correctly reproduce the observed frequencies
(\citealt{goupil11}, \citealt{deheuvels10}, \citealt{silva_aguirre13}).
\cite{deheuvels16}, modeling several MS stars, found that $\ov$ increases with the mass. Moreover, gravity (g) mode pulsators, like slowly-pulsating B (SPB) stars, are interesting targets to constrain the additional mixing around convective cores. Indeed, gravity modes probe the inner chemical structure of the star and allow detailed investigation of the convective core extensions. \citeauthor{moravveji15} (\citeyear{moravveji15}, \citeyear{moravveji16}), when modeling SPB stars, found that overshoot was necessary, and they favored diffusive overshooting over a simple extension of the central mixed region.

Post-main-sequence stars are another way to put constraints on the amount of overshooting. Once the central hydrogen is exhausted, nuclear energy production stops, leaving an inert radiative helium core. This core then contracts, heating the surrounding hydrogen layers of the star until shell burning starts. At that moment, the star begins its subgiant phase, and evolves on a nuclear timescale for masses below about $1.5$ solar masses ($\smass$). For stars that are close to the terminal-age main sequence (TAMS), the structure and the evolution remain highly influenced by the properties of the MS convective core. Interestingly, the star begins to exhibit mixed modes at that moment. These modes behave like gravity modes in the internal propagation cavity and pressure modes in the outer one. Thus, they allow us to finely probe the deepest layers of the star, all the while being observable. This and the proximity of the subgiant to the TAMS make the mixed modes of young subgiants valuable data in studying the extension of convective cores.

Another particularity of mixed modes is their very fast evolution, compared to the nuclear evolution timescale of the subgiant phase. Indeed, mixed mode frequencies change dramatically over the course of a few million years. This makes their seismic modeling challenging. Recently, increasing efforts have been made to model subgiants (\citealt{huber19}; \citealt{stokholm19}; \citealt{metcalfe20}; \citealt{deheuvels20}; \citealt{li19}, \citeyear{li20}), driven by both their great physical interest and the sudden increase of seismic data for these stars. 
Most of those works focused on finding the optimal stellar parameters for one or several subgiants. So far, few studies have used subgiants as tools to test stellar physics, mainly due to the challenges of their modeling, as mentioned above. 

\cite{deheuvels11} successfully constrained $\ov$ from a subgiant observed by CoRoT, HD 49385, which exhibits only one g-dominated mode and is therefore very close to the TAMS. They found that either no overshooting, or a model with $\ov = 0.19$ were giving equally good results. In this work, we modeled a young subgiant, KIC10273246, which was observed by \textit{Kepler} over almost 1000 days. That star exhibits two g-dominated modes, which allows us to better constrain its inner structure. We performed a thorough seismic modeling of the star, in order to precisely estimate its stellar parameters and to place constraints on the extension of its MS convective core. 

In Sect.~\ref{probing_potential}, we show the utility of having access to two g-dominated mixed modes in young subgiants. In Sect.~\ref{observational_properties}, we present the surface observables of KIC10273246 and perform a fresh analysis of its oscillation spectrum using the full \textit{Kepler} data set. We then describe, in Sect.~\ref{modeling_method}, the modeling technique that we adopted, which is an improved version of the method developed by \cite{deheuvels11}. Sect.~\ref{results} presents our optimal stellar models and the constraints that were obtained from the extent of the MS convective core for KIC10273246. We discuss these results in Sect.~\ref{discussion}, and Sect.~\ref{conclusion} is dedicated to our conclusions.

\section{Probing potential of mixed modes}
\label{probing_potential}

Just after the main sequence, the oscillation spectra of solar-like pulsators show the presence of mixed modes, which are due to the coupling between the observed p-modes and low radial-order g-modes ($n_g=1,2,3$, $n_g$ being the number of nodes in the g-mode cavity). The frequency spacing between low-order g-modes is large (several times the large separation of p modes), so that only a few are in the observable frequency window during the subgiant phase. Moreover, with $n_g$ being low, the pure g modes that couple to p modes do not follow an asymptotic behavior (as described in \citealt{shibahashi79}, \citealt{tassoul80}). The oscillation spectra of subgiants therefore constrast with those of more evolved stars, which typically have more g-dominated modes than p-dominated modes, and for which $n_g$ is of the order of several tens (e.g. \citealt{mosser12}). 

\begin{figure}
    \centering
    \includegraphics[width=\hsize]{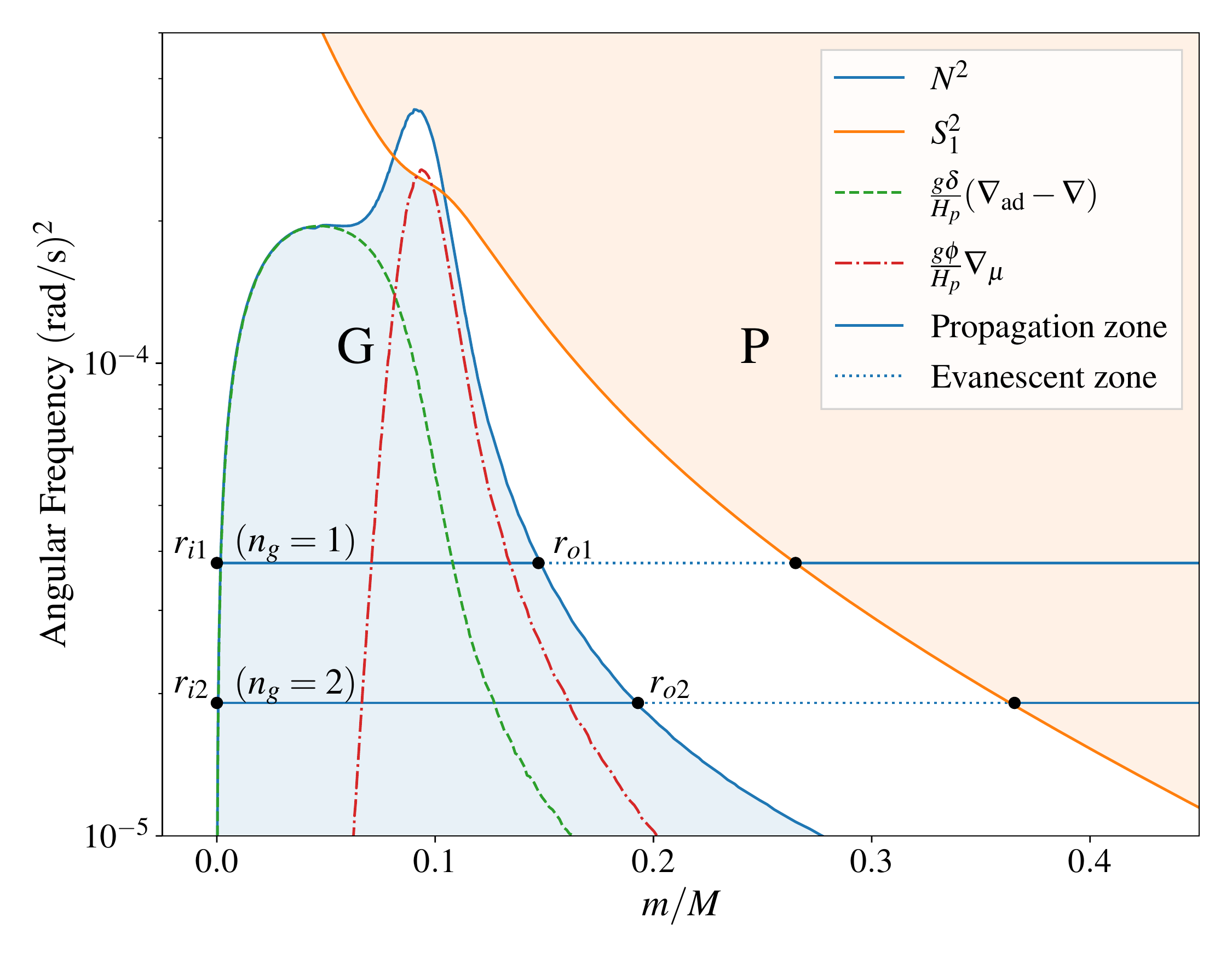}
    \caption{Typical propagation diagram of a low-mass subgiant star. The \N frequency is represented in blue, delimiting the g-mode cavity (light blue). The Lamb frequency, in orange, delimits the p-mode cavity (light orange). Two g-dominated mixed mode angular frequencies, with $n_g=1,2$, are represented (solid lines in propagation zones, dotted lines in evanescent zones). The G cavity turning points are noted as $r_{i1}$, $r_{i2}$ and $r_{o1}$, $r_{o2}$. Finally, the thermal and chemical contributions to the \N frequency are represented in green (dashed) and red (dot-dashed), respectively.}
    \label{propagation_diagram}
\end{figure}

The frequencies of mixed modes are mostly determined by two structural features of the star. The first is the g-mode (G) cavity, which is delimited by the \N frequency $N$. The second is the evanescent zone between the g-mode and p-mode (P) cavities, the latter being delimited by the Lamb frequency $S_l$.

The G cavity affects the frequency of the g-mode that is involved in the mixed mode frequency. G-mode frequencies, in the asymptotic theory, can be approximated by
\begin{equation}
\label{int_gmode}
    \nu_{n,l} \approx \frac{\sqrt{l(l+1)}}{2\pi^2(n-1/2)} \int_{r_1}^{r_2} \frac{N}{r} \dd r,
\end{equation}
$l$ being the degree of the mode, $r_1$ and $r_2$ the turning points in the G cavity, and $r$ the local radius of the star. In our case, $n_g$ is low for the observed modes, so the asymptotic expression given in Eq.~\ref{int_gmode} should not apply. However, it has been shown that it can provide qualitative information about the behavior of the mixed mode frequencies \citep{deheuvels10}. It tells us that g-dominated modes should give strong constraints on the \N frequency in the G cavity. One can write it in the following form (e.g. \citealt{kippenhahn}):
\begin{equation}
    N^2 = \frac{g\delta}{H_p}\left(\nabla_{\mathrm{ad}} - \nabla + \frac{\phi}{\delta}\nabla_{\mu} \right),
\end{equation}
where $g$ is the local gravity, $\delta=-(\partial \ln \rho / \partial \ln T)_{P,\mu}$, $\phi = (\partial \ln \rho / \partial \ln \mu)_{P,T}$, $\nabla_{\mathrm{ad}} = (\partial \ln T / \partial \ln P)_{\rm ad}$, $\nabla = \partial \ln T / \partial \ln P,$ and $\nabla_{\mu}=\partial \ln \mu / \partial \ln P$. The \N frequency consequently carries information about both the thermal (two first terms in parentheses) and compositional structure (last term) of the star. 

The evanescent zone affects the coupling 
between the two cavities, whose strength is linked to the size of this region and the value of $N$ inside it (e.g., \citealt{unno89}).
Using a toy model, \cite{deheuvels11} showed that a strong coupling induces a shift of the $l\geq1$ p-dominated frequencies that are close to a g-mode. The p-dominated frequencies therefore provide complementary information about the internal structure of the subgiant.

In this work, we investigated whether having several g-dominated modes in the observed oscillation spectrum could offer more information regarding the extension of the MS core.
From above, we know that the frequencies of the g-dominated mixed modes are related to the $N/r$ integral between the turning points of the G cavity. Fig.~\ref{propagation_diagram} shows the propagation diagram of a subgiant close to the TAMS, highlighting the frequencies of the first two g-dominated $l=1$ mixed modes, that is, those that arise due to the coupling of p modes with g modes of radial orders $n_g=1$ and 2. We denote their turning points in the G cavity as $r_{i1}$, $r_{o1}$ for the mode with $n_g=1$, and $r_{i2}$, $r_{o2}$ for the mode with $n_g=2$. The difference between the two frequencies is thus mainly related to the \N frequency value between $r_{o1}$ and $r_{o2}$ (as the one in the $[r_{i1}, r_{i2}]$ region is negligible). This region, as it can be seen in Fig.~\ref{propagation_diagram}, is dominated by the $\mu$-gradient contribution. This gradient is related to the characteristics of the hydrogen-burning shell, especially the nuclear energy generation, and thus its temperature and composition. 
It has been shown that a H-burning shell structure depends on the MS core features, and especially on the amount of core overshooting. One can see this in Fig. 5 of \citet{deheuvels10b}, which exhibits two \N profiles of stars having the same evolutionary stage and position in the HR diagram, but computed with different $\ov$. The two profiles differ mainly in the peak caused by the $\mu$-gradient, and these structural differences are large enough to have a significant impact on the eigenfrequencies of the star.

For all those reasons, stars with two visible g-dominated modes are therefore expected to be interesting targets on which to place constraints on the efficiency of core overshooting.
That criterion led us to the choice of KIC10273246, a subgiant with two g-dominated modes and a mass of $1.26 \pm 0.10 \, \smass$ \citep{creevey2012}, which places it safely in the mass range where stars are expected to have a convective core during the MS.

\section{Observational properties of KIC 10273246}
\label{observational_properties}

\subsection{Surface constraints \label{sect_surface}}

\subsubsection{Constraints from spectroscopy}

Surface constraints were obtained for KIC10273246 by \citet{creevey2012}. The authors used different algorithms on the same spectra obtained with the FIES spectrograph.
For our target, they found effective temperatures ($\teff$) ranging from 5933 to 6380 K. A weighted mean gives us a value of $6150 \pm 100$ K, which we have used to constrain our seismic modeling. The star was also found to have to have a sub-solar metallicity, with $\mathrm{[Fe/H]} = -0.13 \pm 0.1\,\mathrm{dex}$. 

\subsubsection{Constraints from broadband photometry}
\label{part_SED}

\begin{figure}
    \centering
    \includegraphics[width=\hsize]{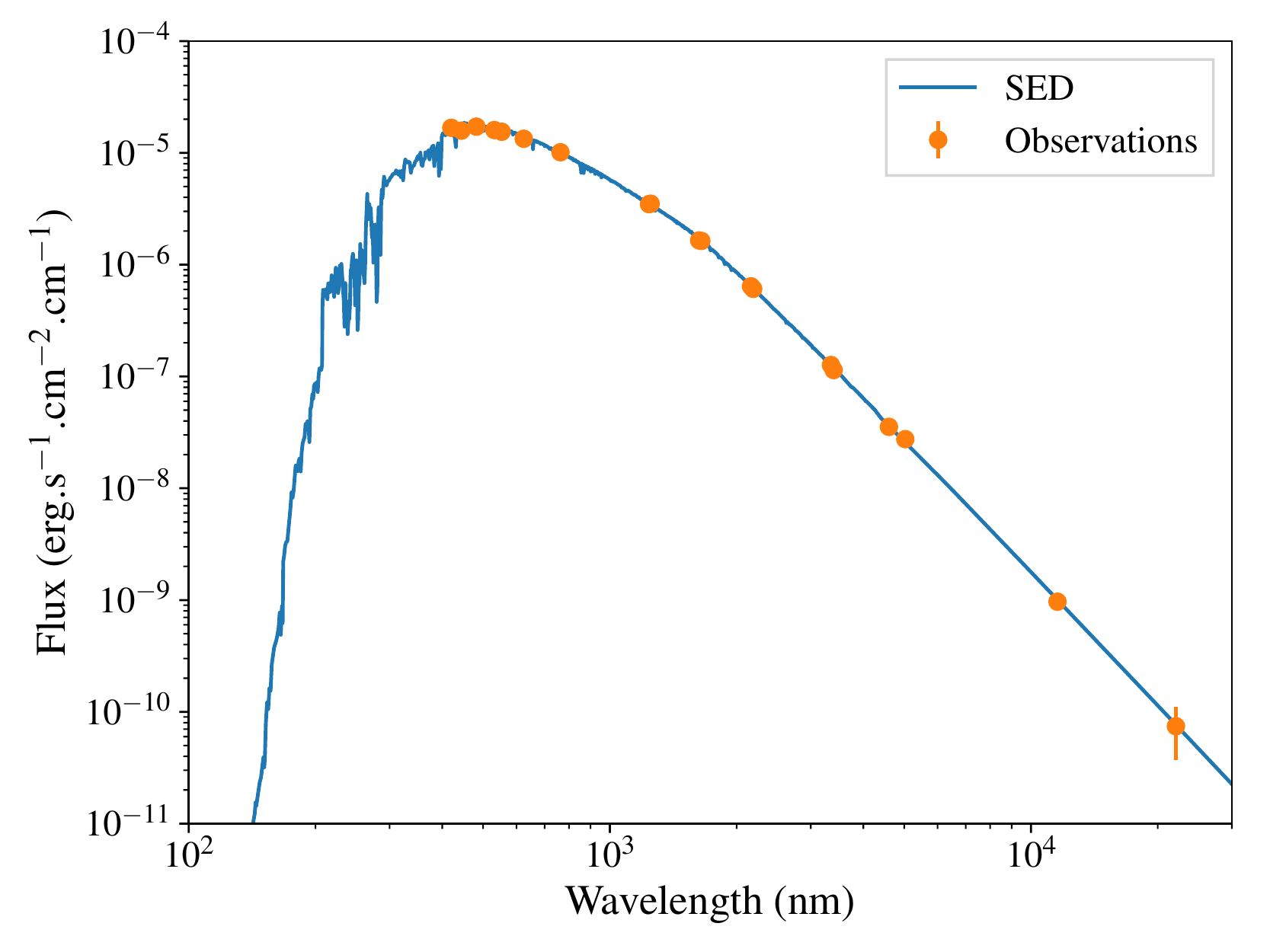}
    \caption{Best fit of the SED using Kurucz atmosphere models. The orange points represent the observations with the corresponding error bars, and the blue curve represents the best fit SED model.}
    \label{fit_SED}
\end{figure}

To obtain a reliable value of the luminosity of the star, we performed a spectral energy distribution (SED) fit, following the procedure of \citet{stassun2016}.
We extracted photometric values using the VizieR photometry tool. Those data come from the NUV filter from \emph{GALEX} \citep{galex}, the $B_T$ and $V_T$ filters from \emph{Tycho-2} \citep{tycho2}, the $J$,$H$ and $K_s$ filters from \emph{2MASS} \citep{2MASS}, the \emph{gri} filters from \emph{SDSS} \citep{2MASS}, the W1-W4 filters from \emph{WISE} \citep{WISE}, and the G magnitude from \emph{Gaia} \citep{gaia}. The atmosphere model comes from the Kurucz atmosphere grid \citep{kurucz2005}, with the surface gravity ($\log g$) derived from the seismic relations (see~Sect. \ref{scaling}), and the metallicity coming from spectroscopic measurements. We then fit the photometry points to the spectrum, with the $\teff$ and extinction $A_v$ as free parameters. We also used the spectroscopic data from \cite{creevey2012} and the extinction from \citet{green2019} as priors. With a reduced $\chi^2$ of 0.7, we found $\teff = 6000 \pm 33 \, \mathrm{K}$, and $A_v = 0.00^{+0.037}_{-0.000} \, \mathrm{mag}$. The fit spectrum and the photometric data are represented in Fig.~\ref{fit_SED}. Finally, we integrated the flux over all the wavelengths and used the distance from \emph{Gaia} to obtain the luminosity of the star. According to \citet{zinn2019}, a parallax bias exists in the Kepler field, which depends on the G-band magnitude and the pseudo-color $\nu_{\mathrm{eff}}$ (effective wavenumber of the photon flux distribution in the Gaia band) of the star. We found $\varpi - \varpi_{\mathrm{Gaia}} = 39.15 \pm 9.46 \, \mu\mathrm{as}$, which gives $L = 5.74 \pm 0.17 \, \lsol$. This result is, as expected, lower than the \emph{Gaia} archive value ($5.92 \pm 0.13 \, \lsol$) due to the parallax offset.

\subsection{Seismic constraints} 

\subsubsection{Preparation of \emph{Kepler} light curve}

The subgiant KIC10273246 was observed with \kepler\ between quarters Q0 and Q11 (total duration of 978 days) in short cadence (58.85 s). An early seismic analysis of the target was performed by \cite{campante2011} using the first four quarters of \kepler\ observations (325 days of data). They estimated the frequencies of oscillation modes of degrees $l=0,1,$ and 2 over eight overtones. We revisited this analysis using the complete \kepler\ data set. The light curve of the star was processed using the \kepler\ pipeline developed by \cite{jenkins2010}. Corrections from outliers, occasional drifts and jumps were performed following the method of \cite{garcia2011}. The power density spectrum (PSD) was then obtained by applying the Lomb-Scargle periodogram (\citealt{lomb1976}, \citealt{scargle1982}). 

\begin{figure}
    \centering
    \includegraphics[width=\hsize]{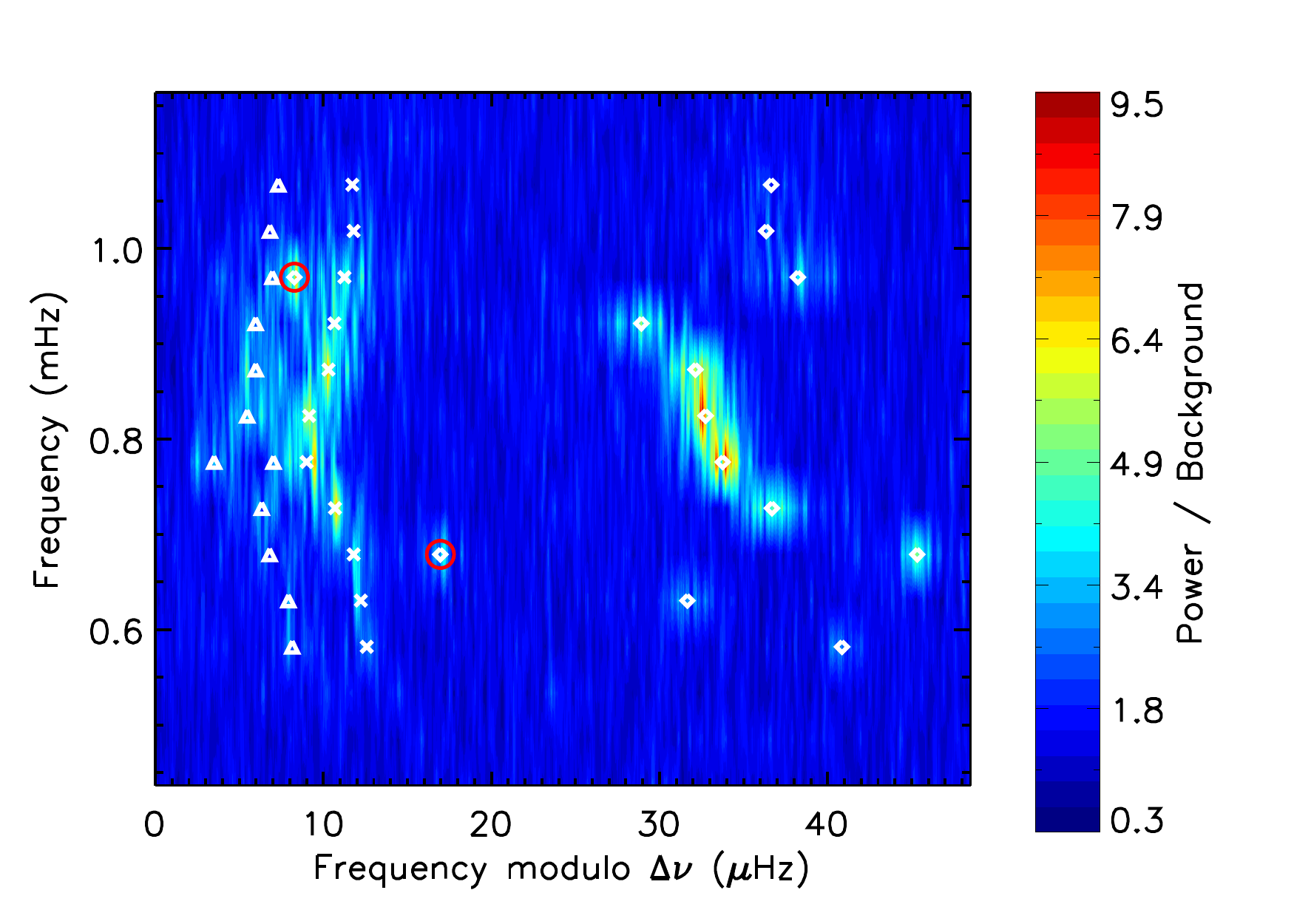}
    \caption{\'Echelle diagram of KIC10273246, folded with $\Delta\nu = 48.2\,\mu$Hz. For clarity, the spectrum was smoothed over a 0.2-$\mu$Hz boxcar. The white symbols indicate the frequencies that have been extracted for modes of degree $l=0$ (crosses), $l=1$ (diamonds), and $l=2$ (triangles) in Sect. \ref{sect_modefit}. The two $l=1$ g-dominated mixed modes are circled in red.}
    \label{fig_echelle_obs}
\end{figure}

The PSD is shown in the shape of an \'echelle diagram in Fig.~\ref{fig_echelle_obs}. We recall that the \'echelle diagram is built by dividing the PSD in consecutive chunks with a length corresponding to the large separation of acoustic modes $\Delta\nu$, and piling them up. Here, we used the estimate of $\Delta\nu = 48.2\,\mu$Hz obtained by \cite{campante2011}. The main interest of \'echelle diagrams is that acoustic modes of the same degree align in nearly straight ridges, which eases mode identification. The neighboring $l=0$ and $l=2$ ridges are readily identified on the left part of the \'echelle diagram (modes indicated by crosses and triangles, respectively, in Fig. \ref{fig_echelle_obs}). 

The ridge of $l=1$ modes (indicated by diamonds in Fig.~\ref{fig_echelle_obs}) deviates from the nearly-vertical  line that is expected for purely acoustic modes. This behavior is known to arise for dipolar modes in the presence of avoided crossings between low-order g modes and p modes in subgiants. Each avoided crossing is characterized by the presence of an additional mode, which lies away from the ridge of theoretical purely acoustic $l=1$ modes (which would be a nearly vertical line at an abscissa of about 35 $\mu$Hz in Fig.~\ref{fig_echelle_obs}). This mode is most strongly trapped in the core and is thus g-dominated. The neighboring $l=1$ modes are p-dominated, but their frequencies are nevertheless affected by the presence of the g-dominated mode. The modes with frequencies larger than the g-dominated mode are shifted to higher frequencies (to the right in the \'echelle diagram) and those with frequencies below the g-dominated mode are shifted to lower frequencies (to the left in the \'echelle diagram). These features are clearly visible in Fig. \ref{fig_echelle_obs}, corresponding to two $l=1$ avoided crossings. The $l=1$ g-dominated modes associated to these avoided crossings are circled in red in Fig. \ref{fig_echelle_obs}. 

\subsubsection{Extraction of oscillation mode parameters \label{sect_modefit}}

To extract the parameters of the oscillation modes, we followed the method of \cite{appourchaux2008}. Here, we briefly recall the main steps of the procedure and refer the reader to that paper for more details.

Prior to fitting the individual oscillation modes, we modeled the background of the PSD. The contribution from granulation was modeled by two Harvey-like profiles, following the prescription of \cite{karoff2013}, and we added a white noise component to account for photon noise. The overall contribution from the oscillations was modeled as a Gaussian function. We fit this model to the PSD using maximum-likelihood estimation (MLE). The central frequency of the Gaussian function gives an estimate of the frequency of maximum power of the oscillations $\nu_{\max}$. To determine the error on this quantity, we subdivided the \kepler\ light curve in ten chunks of equal duration, and fit the background model on the PSD calculated with these time series. The error on $\nu_{\rm max}$ was taken as the standard deviation of the measurements of this quantity for each chunk. We thus obtained $\nu_{\rm max} = 843 \pm 20\,\mu$Hz. The PSD was then divided by the optimal background model.

We then performed a fit of the oscillation modes, which were modeled as Lorentzian functions to account for their finite lifetimes. Each mode profile of degree $l$, radial order $n$, and azimuthal order $m$ was characterized by its central frequency $\nu_{n,l,m}$, its height $H_{n,l,m}$ and its line width $\Gamma_{n,l,m}$. Since dipolar modes have a mixed character, it cannot be assumed that they share similar heights and line widths with the neighboring radial modes, as is often done for main sequence solar-like pulsators. Most quadrupolar modes are expected to be p-dominated, owing to the weak coupling between the P and G cavities for these modes. We therefore assumed that the $l=2$ modes have the same heights and widths as their closest $l=0$ modes, with the exception of one g-dominated $l=2$ mode, which is discussed below and in Sect.~\ref{sect_mode_l2}.
Non-radial modes are split into multiplets by rotation. Owing to the slow rotation of subgiants, the effects of rotation on the mode frequencies can be found by applying a first-order perturbation analysis. The components of a rotational multiplet are thus expected to be equally spaced by the rotational splittings $\delta\nu_{n,l}$. We also assumed that they share similar line widths, and that their height ratios depend only on the inclination angle $i$ of the star following the expressions given by \cite{gizon2003}. In principle, mixed modes can have different rotational splittings, because they probe the rotation at different depths in the star. This has been used to probe the internal rotation of subgiants (e.g., \citealt{deheuvels2014}).

To test whether individual rotational splittings can be measured in KIC10273246, we first performed local fits of the non-radial modes. Following the method described by \cite{deheuvels15}, we fit each mode using two different models: one under the $H_0$ hypothesis (no rotation, so that each mode is modeled as a single Lorentzian), and one under the $H_1$ hypothesis (rotation is considered and each mode is modeled as a set of $2l+1$ Lorentzians separated by the rotational splitting). It is clear that hypothesis $H_1$ necessarily provides better fits to the data than hypothesis $H_0$ since it involves two additional free parameters (inclination angle and rotational splitting). The significance of hypothesis $H_1$ can be tested using the likelihoods $\ell_0$ and $\ell_1$ of the best fits obtained under the $H_0$ and $H_1$ hypotheses, respectively. As shown by \cite{wilks38}, the quantity $\Delta\Lambda\equiv 2(\ln\ell_1 - \ln\ell_0)$ follows the distribution of a $\chi^2$ with $\Delta n$ degrees of freedom, where $\Delta n$ is the difference between the number of free parameters involved in hypotheses $H_1$ and $H_0$ (here, $\Delta n=2$).\footnote{We note that the definition of $\Delta\Lambda$ in Sect. 3.1 of \cite{deheuvels15} contains an erroneous minus sign. This is just a typo and the results presented in the paper consider the correct expression for $\Delta\Lambda$.} For each multiplet, we thus obtained a value of $\Delta\Lambda$. The false-alarm probability was then given by the $p$-value  $p=P(\chi^2(2 \hbox{ dof}) \geqslant \Delta\Lambda)$, which corresponds to the probability that a mode under the null hypothesis can produce such a high value of $\Delta\Lambda$. 

\begin{figure}
    \centering
    \includegraphics[width=\hsize]{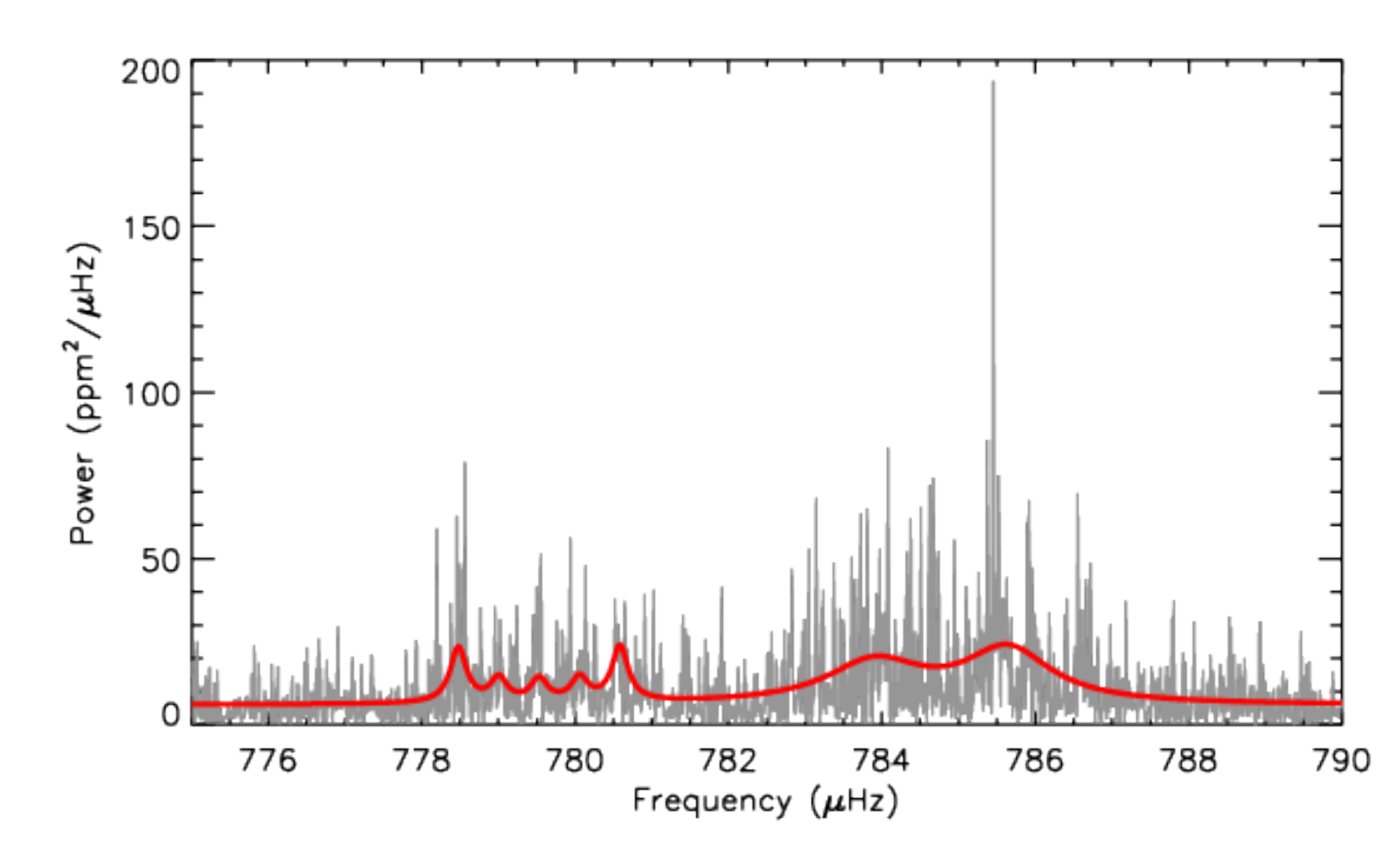}
    \caption{Oscillation spectrum of KIC102773246 in the vicinity of a quadrupolar mode that was found to be significantly split by rotation (see Sect. \ref{sect_modefit}). The thick red curve corresponds to our best-fit model of the spectrum. Two quadrupolar mixed modes are visible (around 779.4~$\mu$Hz and 783.9~$\mu$Hz) and one radial mode (around 785.6~$\mu$Hz).}
    \label{fig_mode_l2}
\end{figure}

For dipolar modes, the lowest $p$-value that we found is 0.08, which is too high to consider the measurement as significant. This means that we cannot reliably extract individual rotational splittings for dipolar modes in this star. The most likely explanation is that the modes have large line widths compared to the rotational splitting. For quadrupolar modes, only one mode (the one with a frequency around 779.4~$\mu$Hz) was found to have a low $p$-value, of about $4\times10^{-5}$, which shows a very high significance level. A rotational splitting of $0.53\pm0.03\,\mu$Hz was obtained for this mode (see Fig. \ref{fig_mode_l2}). This mode is in fact a g-dominated mixed mode, as we show in Sect. \ref{sect_mode_l2}.

We then performed a global fit of the modes (all the modes are fit simultaneously). Since individual splittings cannot be measured, we assumed a common rotational splitting for all $l=1$ and $l=2$ modes (except for the aforementioned $l=2$ mode around 779.4~$\mu$Hz). Since most non-radial modes are p-dominated, we expect the common rotational splitting to essentially measure the rotation in the envelope. The best fit corresponds to a rotational splitting of $\delta\nu = 0.45\pm0.02\,\mu$Hz for non-radial modes and an inclination angle of $i = 55\pm6^\circ$. As was done for local fits, we also performed an additional fit of the modes without including the effects of rotation (null hypothesis). We could therefore estimate the $p$-value corresponding to the measurement of a mean rotational splitting. We found $p \sim 10^{-4}$, which indicates a high level of significance. Our results are compatible with the estimates of \cite{campante2011}, who had found $i \gtrsim 20^\circ$ for this star, and optimal values of the rotational splitting slightly below 0.5~$\mu$Hz. 

The best-fit parameters for the oscillation modes (frequencies, heights, and line widths) are given in Table \ref{tab_seismic_param}. The uncertainties of the fit dipolar mode frequencies range from 0.08 to 0.50~$\mu$Hz. The measured mode frequencies are in quite good agreement with the ones found by \cite{campante2011}. Discrepancies at the level of 3 $\sigma$ were found for only two modes (the dipole mode around 1055 $\mu$Hz and the quadrupole mode around 880 $\mu$Hz). Using the complete \kepler\ data set enabled us to detect $l=0$ and $l=2$ modes  over three additional radial overtones compared to \cite{campante2011}. Our results are also in very good agreement with the recent measurements of mode frequencies for KIC10273246 by \cite{li20} using the complete \kepler\ data set (agreement at the level of 2 $\sigma$ or better for all oscillation modes).

\subsubsection{Detection of an $l=2$ mixed mode}
\label{sect_mode_l2}

We mentioned above that the $l=2$ mode with a frequency of about 779.4~$\mu$Hz is the only mode for which an individual rotational splitting could be measured. This mode also has other distinctive features. It is separated from the closest radial mode by $6.1\pm0.2\,\mu$Hz. By comparison, for the other radial orders, the average separation between the $l=2$ mode and the neighboring $l=0$ mode is 4.4~$\mu$Hz, with a standard deviation of 0.4~$\mu$Hz. This suggests that this mode might be an $l=2$ mixed mode, the frequency of which is modified by the coupling between the p- and g-mode cavities. This hypothesis is strengthened by the fact that it has a short line width ($0.26\pm0.08\,\mu$Hz) compared to the width of the neighboring $l=2$ modes (between 1.7 and 2.4~$\mu$Hz). Indeed, if the mode under study is a g-dominated mixed mode, it should have a higher inertia than p-dominated $l=2$ modes, and therefore a shorter line width. Figure \ref{fig_mode_l2} shows the profile of the radial mode that is closest to the $l=2$ mode under study. There appears to be an additional mode in the left wing of the radial mode, at a frequency of about 783.9 $\mu$Hz. To determine the significance of this mode, we performed local fits assuming either its presence ($H_1$ hypothesis) or absence ($H_0$ hypothesis). We found a $p$-value of 0.01, indicating a reasonable significance level. This also supports the identification of the $l=2$ mode at 779.4~$\mu$Hz as a mixed mode. In this case, the additional mode at 783.9 $\mu$Hz would also be an $l=2$ mixed mode undergoing an avoided crossing with its close neighbor. As is shown in Sect. \ref{results}, the best-fit models for KIC10273246 do show a pair of mixed modes in the vicinity of these two modes, which confirms our identification.

\subsubsection{First estimates of stellar parameters using seismic scaling relations}
\label{scaling}
To obtain first estimates of the global stellar parameters of the star, we used seismic scaling relations, which relate the global seismic parameters $\Delta\nu$ and $\nu_{\max}$ to stellar properties such as the mass, radius and surface gravity \citep{brown91}. These relations could be derived because $\nu_{\max}$ scales to an equally good approximation as the acoustic cut-off frequency (\citealt{brown91}; \citealt{stello08}; \citealt{belkacem11}).

To estimate the asymptotic large separation of acoustic modes, we followed the prescription of \cite{mosser2013}. We fit an expression of the type
\begin{equation}
\nu_{n,0} = \left[ n + \frac{\alpha}{2}\left(n-n_{\rm max}\right)^2 + \varepsilon_{\rm p} \right] \Delta\nu_{\rm obs}
\end{equation}
to the observed radial modes, where $\Delta\nu_{\rm obs}$ is the observed large separation around $\nu_{\rm max}$, $\alpha$ measures the curvature corresponding the to the second-order term in the asymptotic development, $\varepsilon_{\rm p}$ is an offset, and $n_{\rm max} = \nu_{\rm max}/\Delta\nu_{\rm obs}$. We thus obtained $\Delta\nu_{\rm obs} = 48.47\pm0.02\,\mu$Hz, which translates into an asymptotic large separation of $\Delta\nu_{\rm as} = 50.63\pm0.02\,\mu$Hz, following \cite{mosser2013}.

Using our estimates of $\Delta\nu_{\rm as}$, $\nu_{\rm max}$ from Sect. \ref{sect_modefit}, and $T_{\rm eff}$ from Sect. \ref{sect_surface}, we could apply seismic scaling relations to derive preliminary estimates of the star's mass, radius, and surface gravity. We obtained $M = 1.24\pm0.12\,M_\odot$, $R = 2.10\pm0.07\,R_\odot$, and $\log g = 3.88 \pm 0.03$.
\section{Seismic modeling method}
\label{modeling_method}
\subsection{Physics of the models}

We used MESA v10108 \citep{paxton15} evolution models, with OPAL equation of states and opacity tables (\citealt{opal_eos}, \citealt{opal_opacities}), with the solar mixture from \citet{asplund09}. The models were computed with an Eddington-gray atmosphere. The convection regions were treated using the standard mixing length theory (MLT) as prescribed in \citet{cox68}, with a free parameter $\conv$ corresponding to the ratio between the mixing length and the pressure scale height. Microscopic diffusion was taken into account, unless otherwise specified, using the Burgers formalism \citep{bugers69} and diffusion coefficients from \cite{stanton16}. However, radiative accelerations have not been included in the computed models, as the increase in computational time could not be afforded in this study. The impact of those processes are discussed in Sect.~\ref{chemical_diffusion}.

As \citet{gabriel14} stated, for stars that have a growing convective core, it is necessary to use the Ledoux criterion to determine the radius of the convective core $R_{\mathrm{cc}}$. This way, we avoid the creation of unphysical convective zones outside the core in strongly chemically stratified regions, which may have an impact on the composition profile of the star and thus on its evolution. Moreover, we used the predictive mixing scheme \citep{paxton18}.

Core overshooting was modeled as a step extension of the convective core, over a distance
\begin{equation}
\label{definition_alphaov}
    d_{\mathrm{ov}} = \ov \min\left(H_p, R_{\mathrm{cc}} / \conv \right),
\end{equation}
where $d_{\mathrm{ov}}$ is the distance of instant mixing overshooting, $H_p$ the pressure scale height, and $\ov$ a free parameter quantifying the phenomenon. Eq.~\ref{definition_alphaov} replaces the traditional expression $d_{\mathrm{ov}} = \ov H_p$ in order to prevent $d_{\mathrm{ov}}$ from becoming unphysically large when the core is small ($H_p \to \infty$ when $r \to 0$). It is important to note that this definition varies from one evolution code to another (see, e.g., Eq. 1 of \citealt{deheuvels11} for \textsc{Cesam2K}). Low-mass stars have small convective cores, therefore those differences must be kept in mind when comparing models coming from different codes. Additionally, the impact on our results of using a diffusive overshooting, as proposed by \citet{freytag96}, is discussed in Sect.~\ref{diffusive_ov}.

The adiabatic oscillations of the models were computed using ADIPLS \citep{jcd08}, and the surface effects were corrected for using the cubic term of the prescription of \citet{ball14}.

\subsection{Why modeling subgiants is challenging}
\label{difficult_subgiants}
The frequencies of g-dominated mixed modes evolve over a very short time, with a non-linear change of several $\mu$Hz per million years,
which is much larger than the usual uncertainties coming from \textit{Kepler} data. As this timescale is several orders of magnitude shorter than the typical nuclear evolution time of low-mass subgiants, reproducing the mixed modes with a traditional grid technique requires extremely small steps in mass and age. This makes this method prohibitive when the number of free parameters is large, as is required to test the model physics. 
Interpolation in age is possible \citep{li20}, but somewhat difficult for $l=2$ g-dominated modes, which we observed in KIC10273246. Interpolation across tracks (as used e.g., in AIMS, \citealt{rendle19}) could mitigate the need for extremely fine steps in mass, but needs to be tested for subgiants, especially regarding the extreme sensitivity of the g-dominated frequencies to the masses of the models.
Additionally, an ``on-the-fly'' optimization technique may perform badly due to the highly non-linear behavior of the mixed modes, especially during the computation of the derivatives in the case of a gradient-descent kind of algorithm. 

To overcome those difficulties, a new approach is necessary.
We thus developed a nested optimization, where we optimize the physical parameters of models (e.g., metallicity, initial helium abundance etc.) that have been optimized in mass and age beforehand. This way, we can handle those two sensitive parameters using a dedicated and specific procedure, separately from the other ones for which a more traditional technique is possible. This modeling method originates from \citet{deheuvels11} and has been adapted to make it more robust. In the following, we recall the basic principles of this method and highlight the differences with the one used in the present study.

\subsection{Optimization in mass and age}
\label{pre-optim}
\begin{figure}
    \centering
    \includegraphics[width=\hsize]{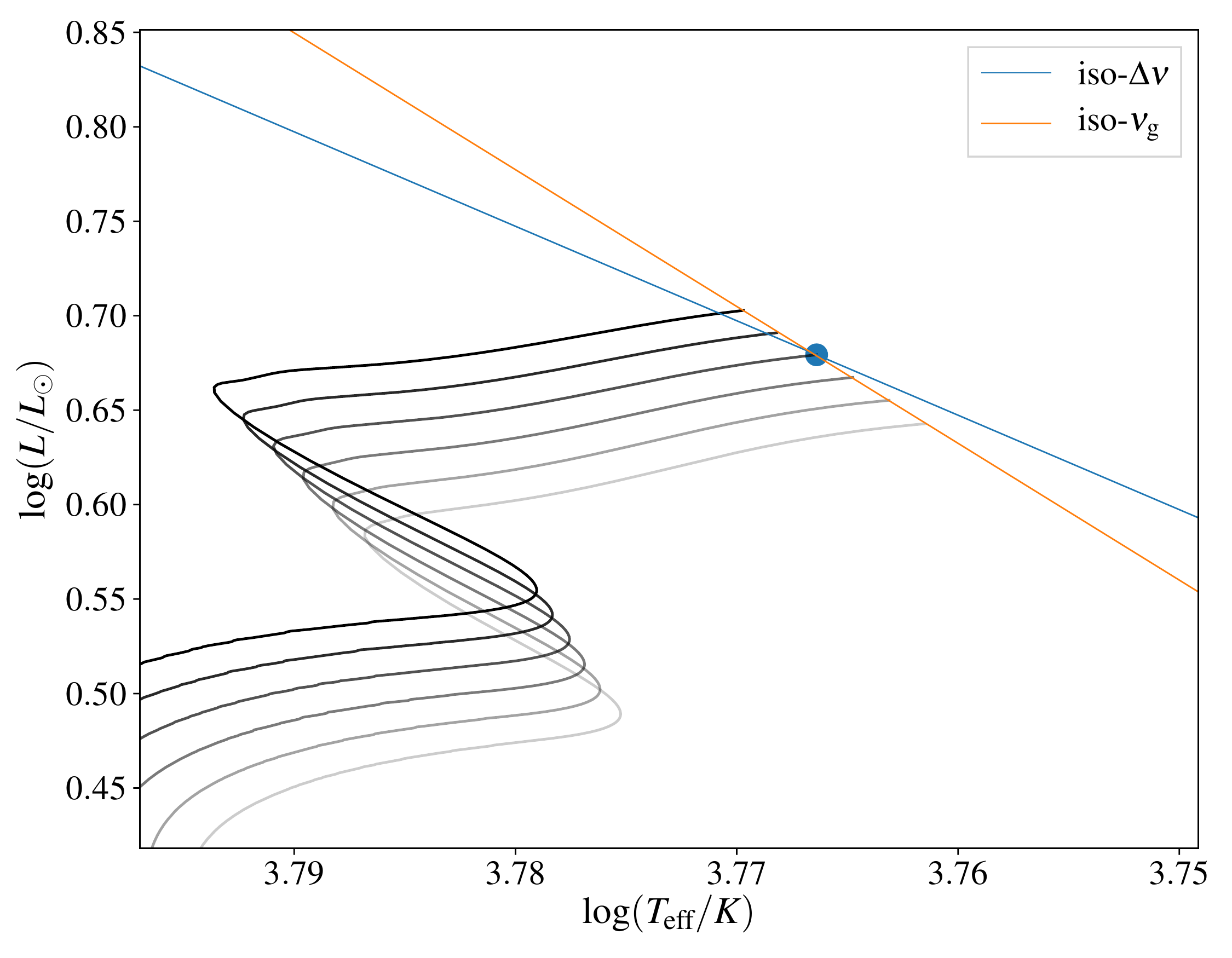}
    \caption{HR-diagram representing the evolution tracks of stellar models with masses varying from $1.2\,\smass$ (lightest gray) to $1.3\,\smass$ (darkest gray) and otherwise identical physics. Each evolution is stopped when $\nug$ is correctly reproduced.}
    \label{HR_optim}
\end{figure}
In that part of the optimization process, we compute models with only two free parameters, the mass and the age of the star, the rest being fixed. The optimization of those two parameters can be made easier thanks to the fact that, if all the other physical parameters (such as metallicity, mixing-length parameter...) are fixed, reproducing only $\Delta \nu$ and the frequency $\nug$ of a g-mode is enough to precisely constrain the mass and the age.

A physical justification of that approach can be found in \citet{deheuvels11}. We remind the reader of it here using a HR-diagram represented in Fig.~\ref{HR_optim}. It shows the iso-$\Delta \nu$ line, as $L \propto \teff^{5}$ for models with the same large separation, and the iso-$\nug$ line, computed from stellar evolution models. The two lines meet at a unique point, that can be reached by tuning only the mass (i.e., choosing the ``right'' evolution path) and age (i.e., stopping at the right moment on that path). In concrete terms, our first step is, at a given mass, to find the age that correctly reproduces the $\nug$ frequency.

\begin{figure}
    \centering
    \includegraphics[width=\hsize]{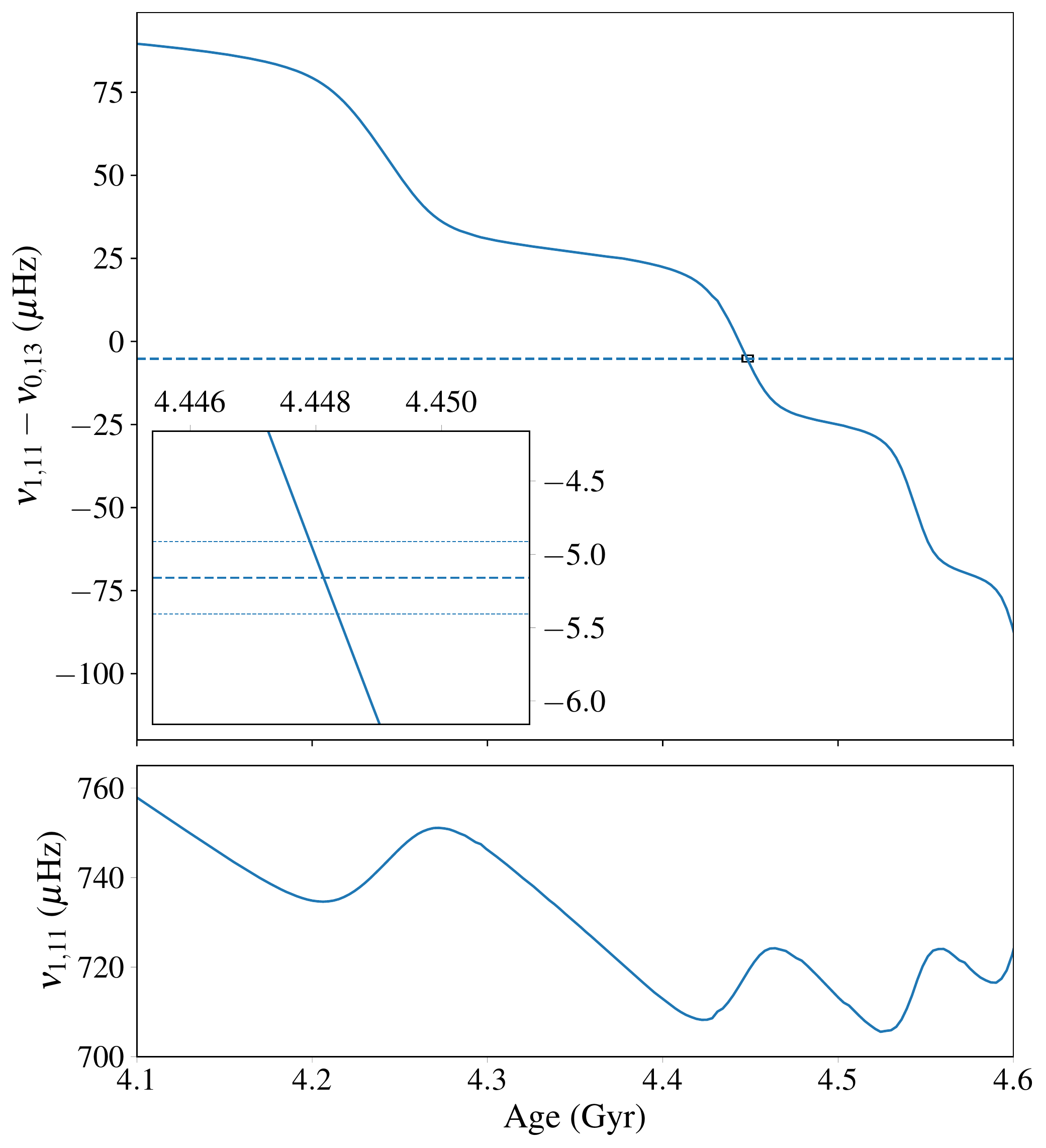}
    \caption{Evolution of $\delta \nu$ (top panel) and $\nu_{1,11}$ (bottom panel) with age for a $1.3\,\smass$ star, after the main sequence. Here, like in our modeling of KIC10273246, $\delta \nu$ is defined as $\nu_{1,11} - \nu_{0,13}$, and the observed value is represented by the dotted line. The plot has been strongly magnified in order to see the 1-$\sigma$ uncertainties from the data. 
    }
    \label{evol_diff}
\end{figure}

As we only see mixed modes and not pure g-modes, we cannot directly measure $\nug$.
A possible solution would be to choose a g-dominated mode (i.e., a non-radial mode far from its ridge) frequency. Unfortunately, such a frequency does not evolve monotonously with age, as can be seen in the upper panel of Fig.~\ref{evol_diff}. We thus preferred to look at the distance between that g-dominated mode and its closest radial mode, which we call $\delta \nu$.

As we can see in the top panel of Fig. \ref{evol_diff}, this value always decreases with age, but it also keeps the interesting properties of the mixed modes as it evolves very quickly during an avoided crossing, allowing us to tightly constrain the age.
This step would be equivalent to following a unique evolution path in Fig.~\ref{HR_optim} and stopping it when it crosses the iso-$\nug$ line.

We then optimize on those ``good-age'' models in order to correctly reproduce the large separation. In practice, to do this we minimize the $\chi^2$ of only the radial modes, which we define as
\begin{equation}
\label{chi2_rad}
    \chi^2_{\mathrm{rad}} = \sum_{\mathrm{n}} \frac{\left(\nu_{\mathrm{0,n}}^{\mathrm{mod}} - \nu_{\mathrm{0,n}}^{\mathrm{obs}}\right)^2}{\sigma_{0,n}^2}.
\end{equation}
We do not take into account the non-radial modes at this stage to eliminate the complexity of behavior of the mixed modes. This approach differs from the one followed by \citet{deheuvels11}, who at this stage searched for models that minimized the difference between the observed average large separation and the one coming from the models. By using all the radial modes instead here, we found that the optimization process is more robust regarding the correction of near-surface effects. It may be observed that the behavior of $\Delta \nu$ (and, in turn, of the radial frequencies) is close to linear when varying the mass. Then, a simple Newton-type algorithm (such as a Levenberg-Marquard algorithm) is enough to quickly find the optimal mass. This step would then be equivalent to the right evolution path that leads to the meeting points of the two iso-lines on Fig.~\ref{HR_optim}.

\begin{figure}
    \centering
    \includegraphics[width=\hsize]{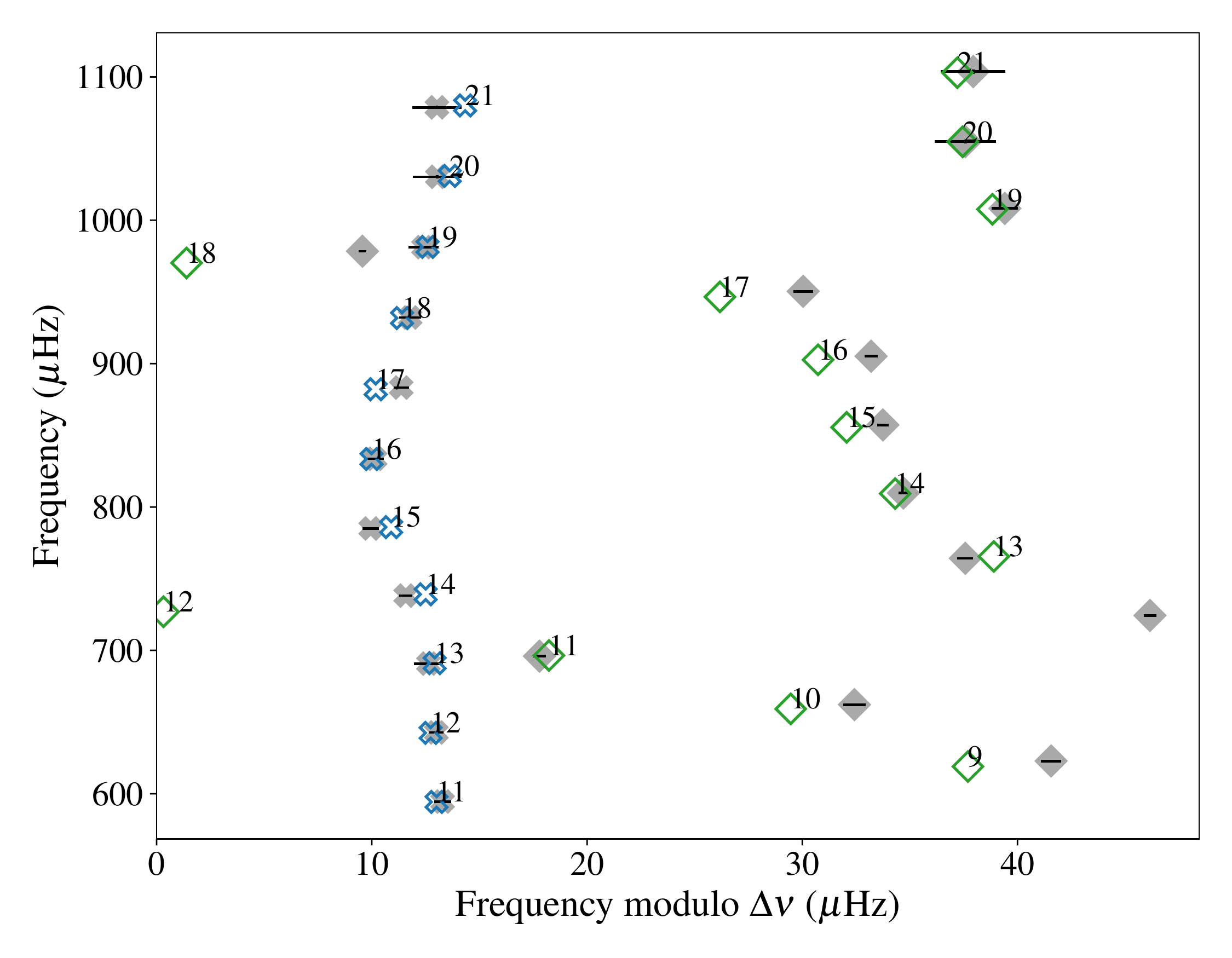}    \caption{\`Echelle diagram of a model optimized in mass and age (open symbols) and of the observed frequencies (full, with their 3-$\sigma$ error bars). The radial and dipolar modes are represented by crosses and diamonds, respectively, with their radial order indicated.  
    }
    \label{pre-opt_echelle}
\end{figure}

Figure \ref{pre-opt_echelle} shows the échelle diagram of a model that we can obtain after that first step, with arbitrary physical parameters: metallicity $\metal = -0.2\, \mathrm{dex}$, mixing-length parameter $\conv = 1.5$, initial helium abundance $Y_0 = 0.28$. We can see that the radial modes and $\delta \nu = \nu_{1,11} - \nu_{0,13}$ (the proxy for $\nug$) are, by construction, correctly reproduced. However, the other frequencies are far from the observed ones. Especially, the g-dominated mode $\nu_{1,18}$ is about 10 $\mu$Hz away from the observations. Thus, to find a better matching model, we adjust the other parameters.

\subsection{Optimizing the other parameters}
\label{optim_glob}

Now that we have a method to correctly find the mass and the age of a star at a given physics, we must find the other parameters of the stars, this time taking into account all the observational constraints. Thus, we define a new $\chi^2$ as
\begin{align}
        \label{chi2_eq}
    \chi^2 
    = \sum_{i=1}^{N_{\mathrm{obs}}} \frac{\left( x_i^{\mathrm{obs}} -  x_i^{\mathrm{mod}} \right)^2}{\sigma_i^2}
    = \sum_{i=1}^{N_{\mathrm{obs}}}  \Delta_i,
\end{align}
where $N_{\mathrm{obs}}$ is the total number of observational constraints, both seismic and non-seismic, $x_i^{\mathrm{obs}}, x_i^{\mathrm{mod}}$  the values of those observed constraints or their computed equivalent, and $\sigma_i$ their respective uncertainties. We also introduced the quantities $\Delta_i \equiv ( x_i^{\mathrm{obs}} -  x_i^{\mathrm{mod}})^2/\sigma_i^2$, which indicate the contributions of every observable to the $\chi^2$, to be used later.

As those parameters have a lower impact on the frequencies than the mass and age, it is now possible to use more traditional approaches. One possibility is to compute grids of models, where each model of the grid is optimized in mass and age. Another option is to perform an optimization using an iterative method, where again each iteration consists of an optimization of the mass and age. To model KIC10273246, we opted for a hybrid method, which is described in the following section.  

\subsection{Fitting procedure adopted for KIC10273246}
\label{modeling_process}

For the modeling of KIC10273246, we left the initial metallicity $[Z/X]_0$, the mixing-length parameter $\conv$, the initial helium abundance $Y_0,$ and, of course, the overshoot parameter $\ov$ as free parameters. At first, to explore the global behavior of the $\chi^2$, we computed a very loose grid ($\metal$ between -0.2 and 0.2, step 0.1; $Y_0$ between 0.24 and 0.28, step 0.02; $\conv$ between 1.5 and 2.1, step 0.2 and $\ov$ between 0.0 and 0.2, step 0.05). We recall that each model of this grid is optimized in mass and age as explained in Sect.~\ref{pre-optim}. The purpose of this loose grid was to investigate whether double solutions or local minima exist. No such features have been found. Moreover, those grids allowed us to identify the region of the best parameters.

We thereafter refined those parameters. As mentioned in Sect.~\ref{optim_glob}, the optimization of [Z/X], $\conv$, $Y_0$, $\ov$ can be performed either through a grid approach or an iterative procedure. We therefore conducted several tests, using stellar models as mock observations, to determine which method is preferable. We found the best robustness when following a hybrid approach: for given values of $Y_0$ (0.26 through 0.31, step 0.01) and $\ov$ (0 through 0.25, step 0.05), we conducted iterative optimizations with the Levenberg-Marquardt algorithm to find optimal values of $\metal$ and $\conv$. This method differs from the one used in \cite{deheuvels11} where a single grid was used for all the free parameters.

Among those models, we considered only those that were compatible with the observational estimates of the chemical enrichment law $\Delta Y_0 / \Delta Z_0$. Typical values quoted for $\Delta Y_0 / \Delta Z_0$ range from 1.4 to 4 (e.g., \citealt{chiappini02}, \citealt{balser06}, \citealt{casagrande06}). We consequently had a conservative approach and took into account all models with $\Delta Y_0 / \Delta Z_0 < 5$.

\section{Results}
\label{results}
In this section, we describe the general characteristics of the best models, before commenting the constraints on $\ov$. We finally investigate at the internal structures of the best models and the constraints brought by the mixed modes.

\subsection{General characteristics of the best models}
\label{general_carac_sec}
\begin{table*}[]
\centering
    \begin{tabular}{c c c c c c c c c c}
    \hline \hline
    $\ov$ & Age (Gyr) & $M/M _{\odot}$ & $R/R_{\odot}$ & $T_{\mathrm{eff}}$ (K)& $L / L_{\odot}$ & $[Z/X]_0$ (dex) & $\conv$ & $Y_0$ & $\chi^2$\\
    \hline
    Uncert. &  0.25 & 0.030 & 0.021 & 83 & 0.20 & 0.010 & 0.089 & 0.020 & -- \\
    \hline
        0.00 & 4.08 & 1.21 & 2.11 & 6109 & 5.60 & 0.005 & 1.77 & 0.29 & 315\\
        0.05 & 3.89 & 1.20 & 2.10 & 6187 & 5.85 & -0.034 & 1.81 & 0.29 & 255\\
        0.10 & 4.03 & 1.22 & 2.11 & 6134 & 5.72 & -0.030 & 1.74 & 0.28 & 201\\
        0.15 & 3.88 & 1.22 & 2.11 & 6192 & 5.89 & -0.073 & 1.74 & 0.28 & 127\\
        0.20 & 3.96 & 1.27 & 2.12 & 6226 & 6.11 & -0.155 & 1.64 & 0.24 & 446\\
        0.25 & 3.26 & 1.31 & 2.13 & 6537 & 7.50 & -0.184 & 2.06 & 0.26 & 3020\\

    \hline
    \end{tabular}
    \caption{Characteristics of the best models, for every value of $\ov$.\label{charac_models}}
\end{table*}

\begin{figure}
    \centering
    \includegraphics[width=\hsize]{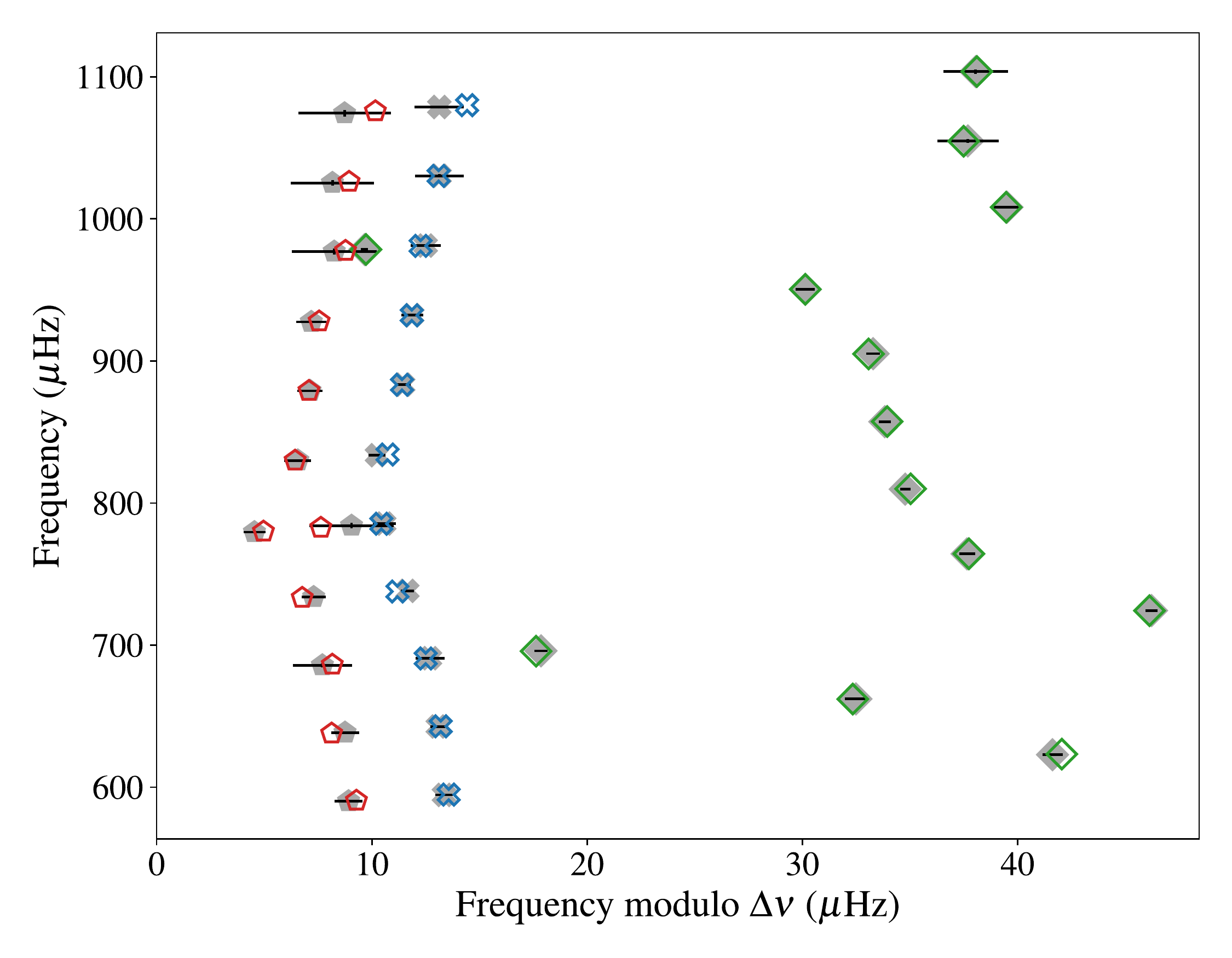}
    \caption{Echelle diagram of the best model, with $\ov=0.15$. Symbols and colors are identical to those in Fig.~\ref{pre-opt_echelle}.}
    \label{echelle_best}
\end{figure}

Following the method described in Sect.~\ref{modeling_process}, we obtained optimal models for each value of $\ov$, whose characteristics are in Table~\ref{charac_models}. The best model, with $\ov=0.15$, has a reduced $\chi^2$ of 3.2. The échelle diagram of this model is represented in Fig.~\ref{echelle_best}. Also, surface observables are consistent with the ones found in the literature, or with the SED fitting previously described: we find a less than 1-$\sigma$ difference for the effective temperature $\teff$, the metallicity $\metal,$ and the luminosity $L$. We found a radius and a mass that are consistent with the seismic scaling relations as well. We can note that, as expected from the mass-based prediction, all good models had a convective core during the MS. This supports our choice of using this star to constrain $\ov$.

We note that our best-fit models are significantly less massive and are older than those of \citet{li20}, who also performed a seismic modeling of KIC20173246 and found $M=1.49\pm0.08\,\smass$ and an age of $2.84\pm0.60\,\mathrm{Gyr}$. These discrepancies could be partially explained by the different assumptions made on the input physics. For instance, \citet{li20} considered a solar-calibrated mixing length parameter, while we left this parameter free in our study. Also, contrary to us, \citet{li20} neglected element diffusion and adopted the mixture of \citet{gs98}. Finally, we stress that the agreement with the observed dipole mode frequencies, in particular for the g-dominated mode $\nu_{1,18}$, is significantly better in the present study than it is for the best-fit models of \citet{li20} (compare Fig. 10 of \citealt{li20} to Fig. 8 of the present paper). These mismatches between models and observations for dipole modes are acknowledged by \citet{li20}, and the authors attribute them to an imprecise modeling of the core structure.

For each combination of ($\ov$, $Y_0$), our minimization using the LM algorithm can be used to derive uncertainties in the stellar parameters. The error bars in the free parameters of the fit ($\metal$ and $\conv$) are obtained as the diagonal coefficients of the inverse of the Hessian matrix. The uncertainties on the other parameters can then be obtained using Eq. 10 of \cite{deheuvels16}. We thus obtain very small error bars, of the order of 0.007 for $\metal$ and 0.002 for $\conv$, which translates into uncertainties of approximately $0.004\,\smass$ for the stellar mass and $0.04\,$Gyr for the age. This means that for a given combination of ($\ov$, $Y_0$), the available observables provide very strong constraints on the stellar parameters. By comparison, we find that optimal models with different $Y_0$ can yield similar agreement with the observations (statistically equivalent $\chi^2$) but have quite different stellar parameters. This degeneracy of stellar models with respect to $Y_0$ is addressed in more detail in Sect.~\ref{Y0_degen}. It thus seems that the uncertainties in the stellar parameters are dominated by the model degeneracy in $Y_0$. We thus used the optimal models with different $Y_0$ to estimate uncertainties in the stellar parameters. For a given $\ov$, we fit a second order polynomial to the $\chi^2$ curve and retrieved the interval of values corresponding to $\chi^2_{\min} + 1$. This gave us the 1-$\sigma$ uncertainties, which are reported in Table~\ref{charac_models}.

\subsection{Constraints on core overshooting}
\label{with_overshooting}

\begin{figure}
    \centering
    \includegraphics[width=\hsize]{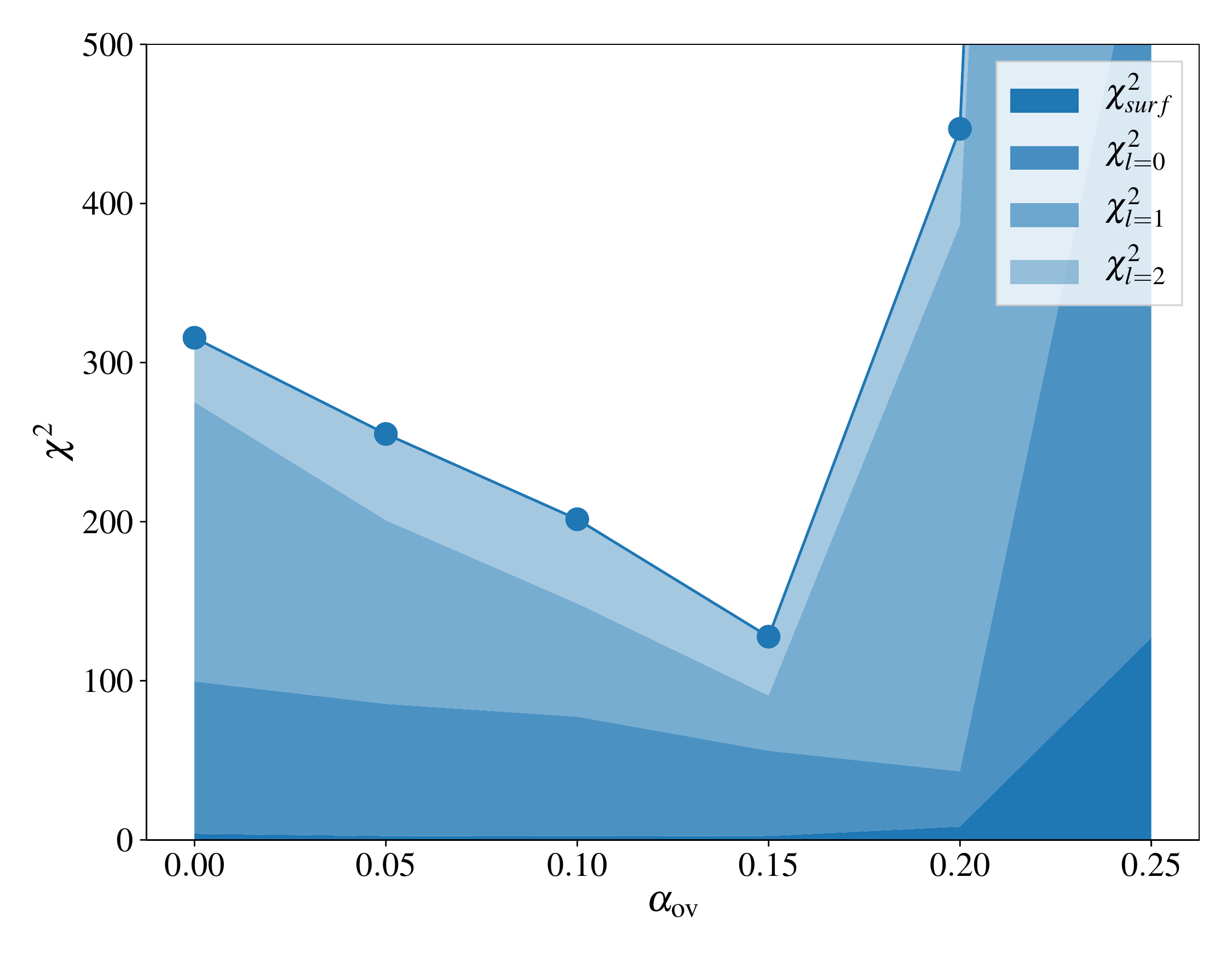}
    \caption{$\chi^2$ of the best model for every value of overshooting. The colored regions indicate the contributions to the $\chi^2$ (i.e., the sum of $\Delta_i$) of the surface observables ($\teff$, $L$ and $[\mathrm{Fe}/\mathrm{H}]$) and the frequencies according to their degrees.}
    \label{chi2_ov}
\end{figure}
\begin{figure}
    \centering
    \includegraphics[width=\hsize]{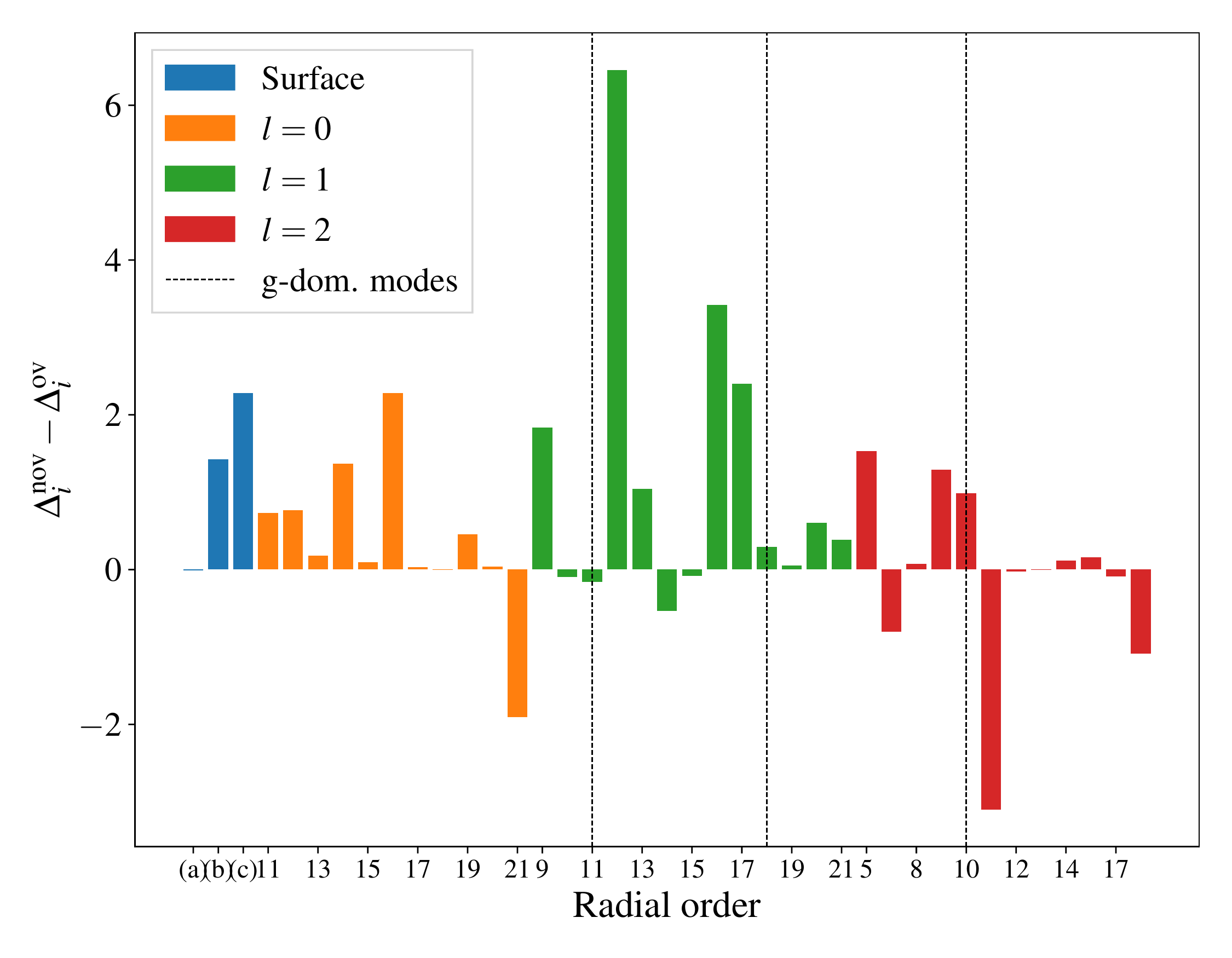}
    \caption{Difference of the $\chi^2$ contributions of the different observables, between $\ov=0.0$ and $0.15$ models. The two dipolar g-dominated modes are represented by dotted vertical lines. (a), (b), and (c) are $\teff$, $L,$ and [Fe/H], respectively.}
    \label{chi2_diff}
\end{figure}

Figure \ref{chi2_ov} shows the variation in the $\chi^2$ of the optimal models with $\ov$. We can see that adding overshooting allows us to reproduce the observed frequencies significantly
better, with the $\chi^2$ of the models without overshoot and with $\ov=0.15$ being 315 and 127, respectively.

To better understand which frequencies allow us to favor models with overshoot, we investigated the contributions of the different observables to the $\chi^2$ ($\Delta_i$ in Eq.~\ref{chi2_eq}). We denote the $\chi^2$ contributions of the observables coming from the optimal models without overshoot as $\Delta^{\mathrm{nov}}_i$ and the equivalent from models with $\ov = 0.15$ as $\Delta^{\mathrm{ov}}_i$. Figure \ref{chi2_diff} represents the differences $\Delta_i^{\mathrm{nov}} - \Delta^{\mathrm{ov}}_i$, positive values meaning that the observable is better reproduced by the $\ov=0.15$ model. As expected, we can see that the main $\chi^2$ difference is due to the dipolar modes, which have a mixed behavior. However, we observe that the g-dominated modes (indicated by the dotted vertical lines) hardly contribute to distinguishing the models with and without overshooting. Both types of model fit the g-dominated frequencies well. The main contributors to the $\chi^2$ differences are in fact the $l=1$ p-dominated modes in the neighborhood of the g-dominated modes. As explained in Sect.~\ref{probing_potential}, the frequencies of these modes are mainly influenced by the coupling between the P and the G cavities. The intensity of that coupling thus accounts for the main part of the differences between the models. We note that all those models correctly reproduce $\nu_{1,18}$, as the high sensitivity of this g-dominated mode strongly constrains the region of parameters of the models with the smallest $\chi^2$.
The major role played by the dipolar modes in the constraints on $\ov$ is also illustrated in Fig.~\ref{chi2_ov}, where the colored regions indicate the contributions to the $\chi^2$ of the surface observables and modes depending on their degree.

Moreover, Fig.~\ref{chi2_ov} indicates that the contribution to the $\chi^2$ of the $l=2$ modes hardly changes with $\ov$. This was partly expected, because their evanescent zone is larger than that of the dipole modes, making the coupling between the G and P cavities weaker. Most of the detectable modes are therefore very strongly p-dominated and do not constrain the deep structure of the star, hence $\ov$. Yet, one g-dominated $l=2$ mode was detected ($\nu_{2,10}$, see Sect.~\ref{sect_mode_l2}). It is interesting to see that, in a similar way to the previous paragraph, its frequency is equally well reproduced by models with and without overshooting. One can see this in Fig.~\ref{chi2_diff}, where $\Delta_i^{\mathrm{nov}} - \Delta^{\mathrm{ov}}_i$ of that mode is less than 1-$\sigma$. On the other hand, the (2,11) mode, whose frequency is close enough to $\nu_{2,10}$ to be influenced by the coupling, varies substantially with $\ov$. Figure \ref{chi2_diff} shows a 3-$\sigma$ difference, despite the high $0.65\,\mu$Hz observational uncertainty, confirming the key role of the coupling in the constraint on $\ov$. Interestingly, however, while the $\ov = 0.15$ model better reproduces the dipolar modes, the (2,11) mode is better fit in the model without overshooting. Nevertheless, its large observational uncertainty prevents it from being too constraining.

Finally, we notice that adding a larger amount of overshooting ($\ov > 0.15$) strongly worsens the quality of the fit, placing a strong maximum limit on the value of $\ov$. To better understand this behavior, we investigate the seismic constraints on the internal structure of the models in
the next section.

\subsection{Constraints on the internal structure from mixed modes}

\subsubsection{Constraints on central density}

\begin{figure}
    \centering
    \includegraphics[width=\hsize]{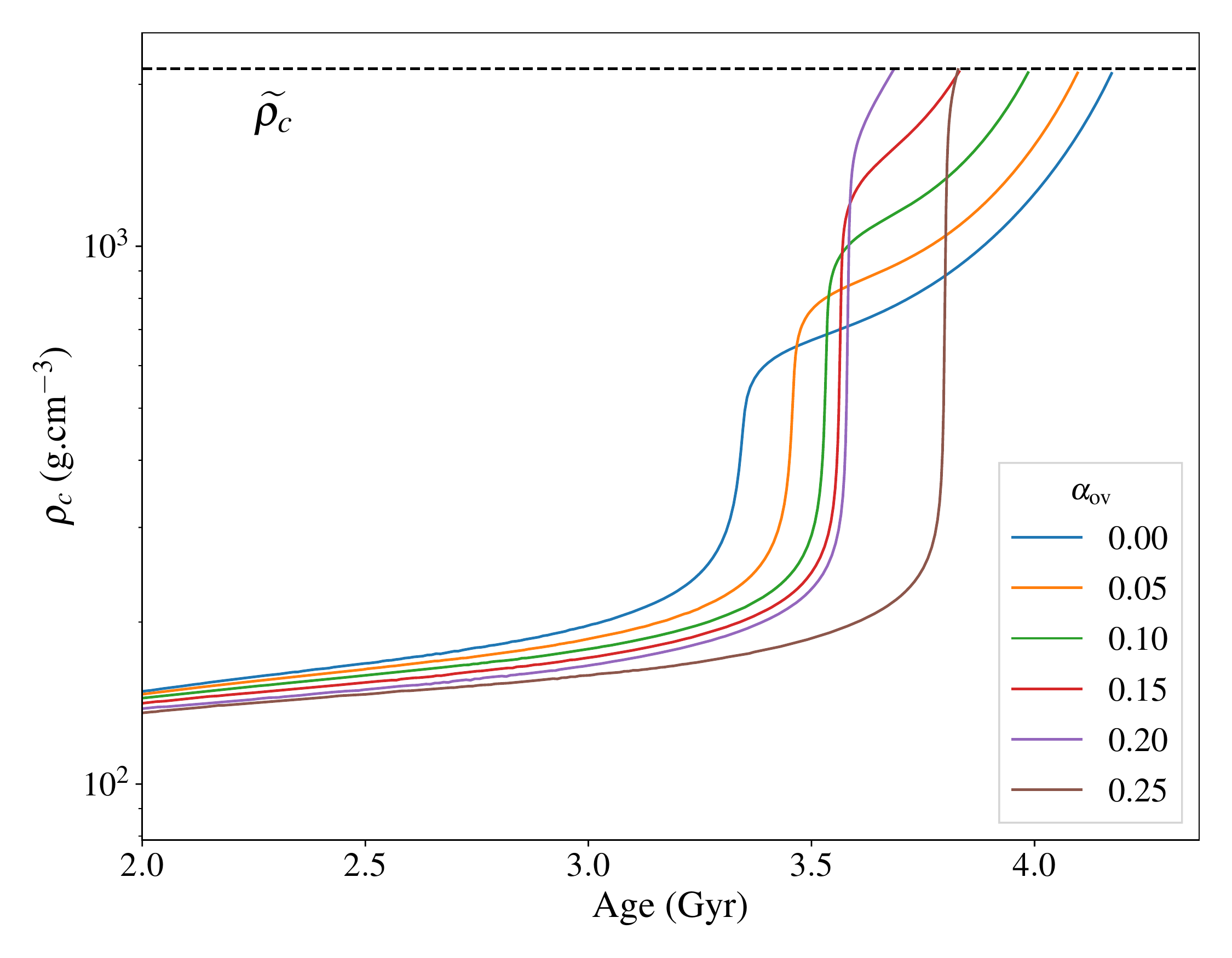}
    \caption{Evolution of $\rho_c$ with age, for models with different $\ov$ that have been optimized in mass and age in order to reproduce $\Delta \nu$ and $\nu_g$.}
    \label{evol_rhoc}
\end{figure}

A tight link is expected between the g-dominated frequencies and the central density $\rho_c$. This comes from the relation between the g-mode frequencies and the \N frequency, which approximately scales as $\rho_c$ (see Eq. 15 from \citealt{deheuvels11}). To verify this, we investigated the constraints placed on $\rho_c$ by the frequency of a g-dominated dipolar mode. For this purpose, we considered the values of $\rho_c$ in the models computed in the loose grids defined in Sect.~\ref{modeling_process}, in which models are optimized in mass and age in order to reproduce $\Delta \nu$ and the frequency of a g-dominated mode (here $\nu_{1,11}$, as described in Sect.~\ref{pre-optim}).
We observed that, despite the wide range of parameters considered, the standard deviation of $\rho_c$ among those models is as low as $32.4\,\mathrm{g.cm}^{-3}$, which represents around 1\% of the mean central density $\widetilde{\rho_c} = 2100\,\mathrm{g.cm}^{-3}$. This standard deviation even decreases to $10\,\mathrm{g.cm}^{-3}$ if we only keep the 200 best models, illustrating the tight relation between $\rho_c$ and the frequency of a g-dominated mixed mode.

This plays a role in the increase of $\chi^2$ for $\ov > 0.15$. To illustrate this point, we computed models with the same values of [Z/X], $Y_0$ , and $\conv$, but with different values of $\ov$. Each of these models was optimized in mass and age, as described above. Fig.~\ref{evol_rhoc} shows the evolution of $\rho_c$ with age for those models. One can see that they all reach approximately the same value of central density, $\widetilde{\rho_c}$, in accordance with the previous paragraph. Moreover, the intensity of the $\rho_c$ jump that is due to the post-MS core contraction increases with $\ov$. We can explain this by the fact that for bigger cores, the layers where hydrogen remains are more distant from the center. They are colder and require a stronger core contraction, hence a stronger jump in $\rho_c$, to reach the fusion temperature. Therefore, when $\ov$ increases, models that reach $\widetilde{\rho_c}$ are closer to the TAMS. When the model gets too close to the TAMS, this impacts the whole stellar structure. Internally, the $\mu$-gradient in the core is very much affected, because the nuclear reactions in the H-burning shell did not have enough time to smooth its shape. In the outer layers, the convective envelope, which expands during the subgiant phase, has a different size. Those processes alter the frequencies (the g- and p-dominated ones, respectively), which are thereby not compatible with the observations.

\subsubsection{Constraints on the \N profile.}

\begin{figure*}
    \centering
    \includegraphics[width=\hsize]{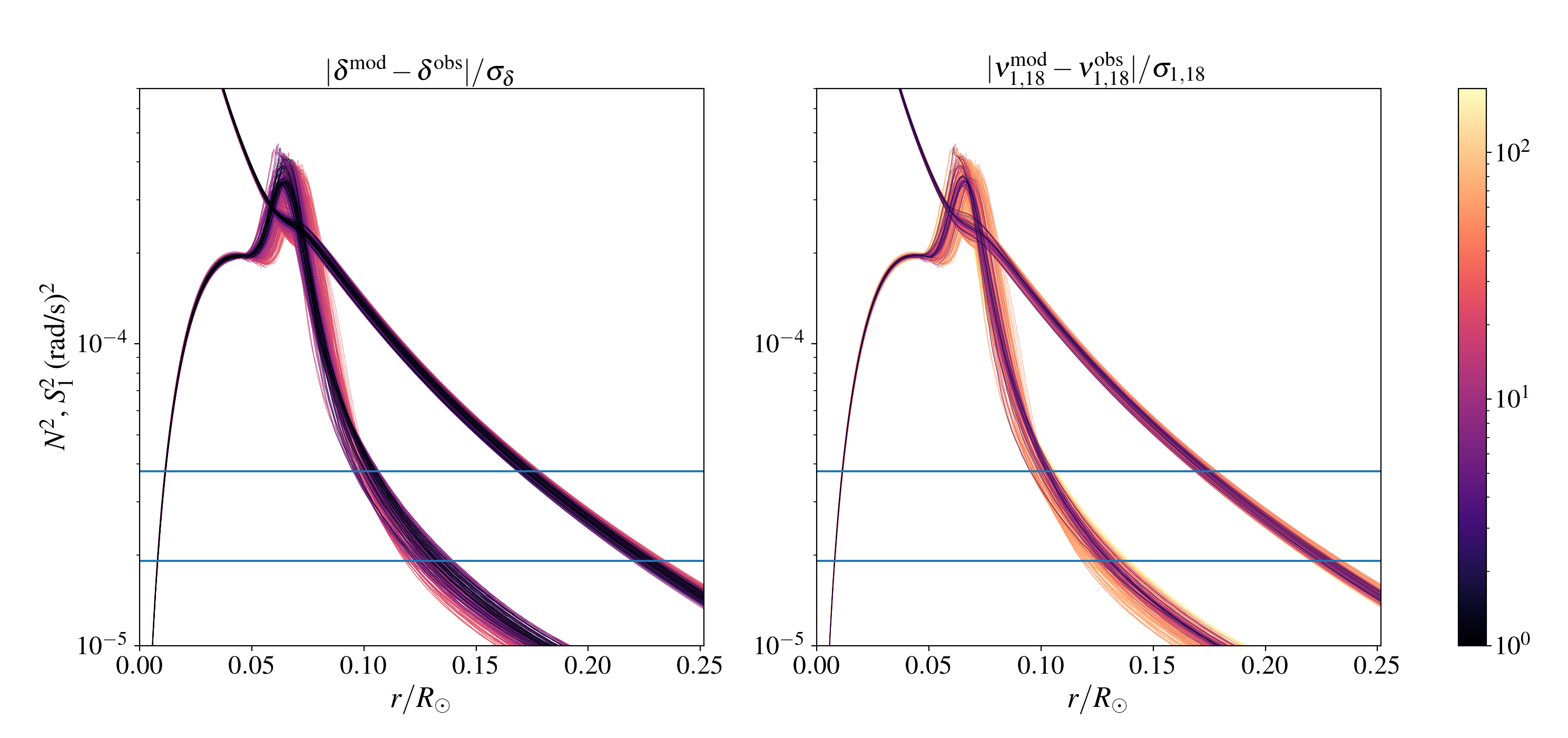}
    \caption{$N^2$ profiles of the loose grid models. The left and right panel profiles are colored in relation to the difference between the models and the observations of $\nu_{1,18}$ and $\delta$, respectively. Those differences are normalized by the observational uncertainties. The blue horizontal lines represent the observed g-dominated frequencies. }
    \label{N2_color}
\end{figure*}

Based on Eq.~\ref{int_gmode}, we expect the frequency of mixed modes to place constraints on the integral of the \N frequency in the G cavity. To investigate this, in Fig.~\ref{N2_color} we plot the $N^2$ profiles for the models of the loose grid defined in Sect.~\ref{modeling_process}, which all reproduce the low-frequency g-dominated mode ($\nu_{1,11}$) and $\Delta \nu$. We observe that reproducing both already strongly constrains the part of the \N frequency dominated by the thermal term ($\nabla_{\mathrm{ad}} - \nabla$), which corresponds to the most central layers ($r < 0.05\, R_{\odot}$). This was expected because of the $1/r$ factor in the Eq.~\ref{int_gmode} integral. On the contrary, the part of $N^2$ that is dominated by the $\mu$-gradient changes significantly within the grid.

We expect that part to be strongly determined by the dipolar modes. We therefore
investigated the constraints brought by the two most determining seismic features (see Sect.~\ref{probing_potential}): the coupling intensity and the frequency of pure g-modes. As those two are not directly measurable, we used observational proxies. The intensity of the coupling can be quantified by $\delta \equiv \nu_{1,12} - \nu_{1,11}$, which is from \citet{deheuvels11}. That value is the difference between the low-frequency g-dominated mode ($\nu_{1,11}$) and the following dipolar p-dominated mode ($\nu_{1,12}$). Thus, $\delta$ increases with the coupling. 
The frequency of a pure g-mode is measured through $\nu_{1,18}$, which is the high-frequency g-dominated mode. For those two values, we color-coded, in Fig.~\ref{N2_color}, the profiles of the \N and Lamb frequencies based on their agreement with the observations (left panel for $\delta$ and right panel for $\nu_{1,18}$).

One can see on the right panel that models correctly reproducing the coupling (i.e., dark profiles) have very similar H-burning shell positions ($N^2$ peak around $r=0.05\, R_{\odot}$). 
However, the \N profiles become more degenerate for higher $r$: several different profiles can reproduce $\delta$ within 1-$\sigma$.
This degeneracy is lifted thanks to the high-frequency g-dominated mode: on the left panel, models closely reproducing $\nu_{1,18}$ all have similar \N profiles. 
This corroborates our theoretical approach in Sect.~\ref{probing_potential}: the high-frequency g-dominated mode adding tight constraints on the shape of the $\mu$-gradient. 
Important gains in structural constraints are therefore obtained from having a second detectable g-dominated mode.

\section{Discussion}
\label{discussion}

\subsection{Diffusive overshooting}
\label{diffusive_ov}
During this study, overshooting was modeled as a step extension of the convective core. However, \citet{freytag96} proposed another prescription, based on results coming from 2D simulations of A-stars and white dwarfs. Overshoot is then modeled as a diffusive process, with a coefficient $D_{\mathrm{ov}}$ exponentially decaying from the boundary of core, following the expression
\begin{equation}
    D_{\mathrm{ov}} = D_{\mathrm{conv}} \exp \left[-\frac{2(r-R_{\mathrm{cc}})}{f_{\mathrm{ov}}H_p} \right],
\end{equation}
with $D_{\mathrm{conv}}$ being the MLT derived coefficient taken just below $R_{\mathrm{cc,}}$ and $f_{\mathrm{ov}}$ a free parameter that tunes the length scale of the overshooting region.

In order to compare the results coming from the two types of overshooting, we first found the $\fov$ that is equivalent to a given value of $\ov$. In order to do this, we searched for the value of $\fov$ that gives models reaching the TAMS at the same age as models computed with a step overshooting and $\ov = 0.15$. We found $\fov = 0.01$. After that, we modeled it using a method similar to the one used in Sect.~\ref{results} and compared the best model with the one computed with a step overshoot. 

\begin{figure}
    \centering
    \includegraphics[width=\hsize]{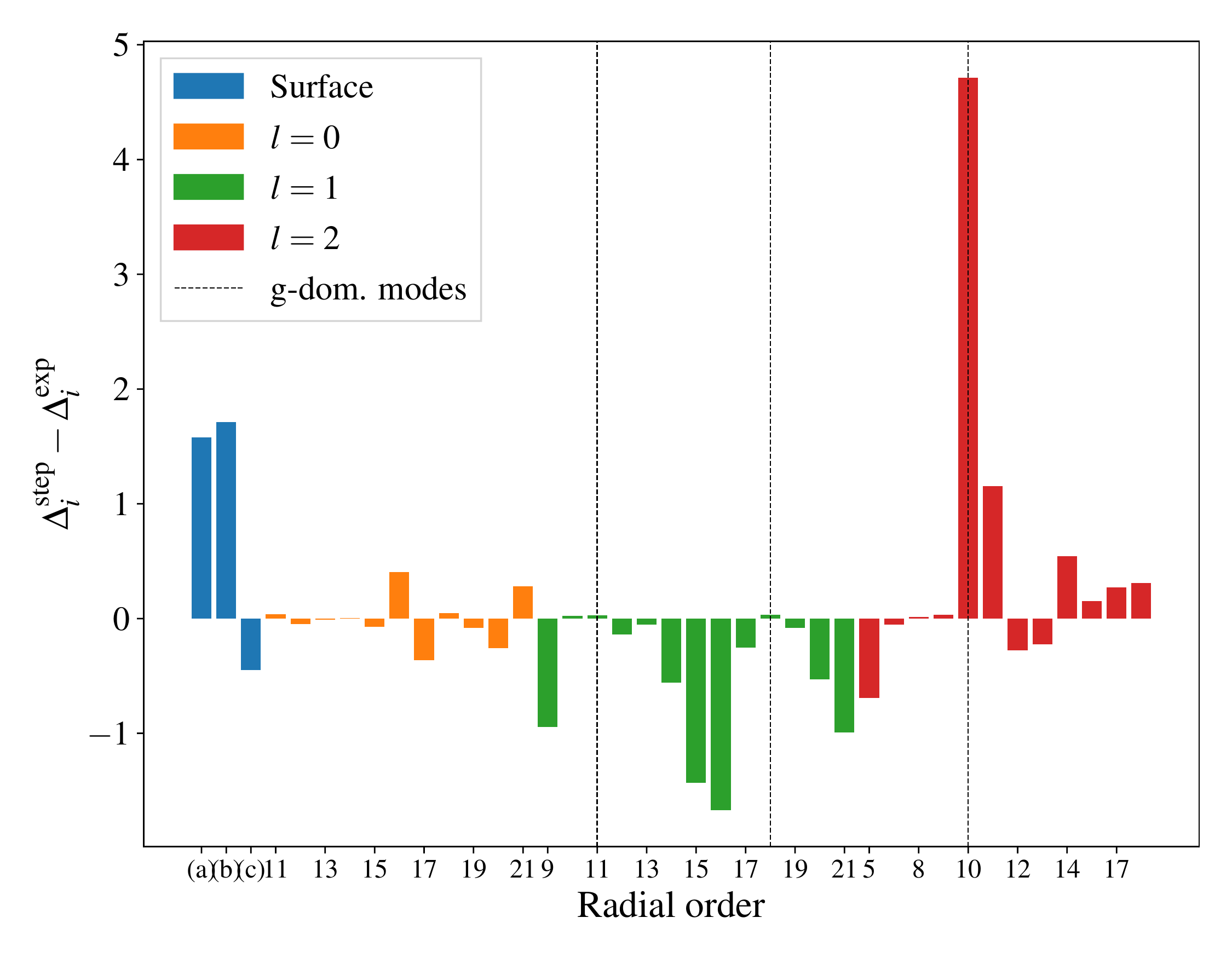}
    \caption{Differences in the $\chi^2$ contributions of the observables coming from the best models with step and diffusive overshoot.}
    \label{diff_freq_exp}
\end{figure}

As we can see in Fig.~\ref{diff_freq_exp}, the differences between the frequencies and observables of the best models with step and diffusive overshoot are mainly less than 1-$\sigma$. We note the exception of the g-dominated $\nu_{2,10}$ mode, which is better reproduced by the diffusive overshoot model. However, its impact on the global $\chi^2$ is counter-balanced by the generally better reproduced dipolar frequencies of the step overshoot model. Moreover, the difference between the characteristics of the two models are within the uncertainties of Table~\ref{charac_models}. Therefore, we cannot discriminate between the two kinds of modelings with the current set of data.

\subsection{Effect of microscopic diffusion}
\label{chemical_diffusion}
\begin{figure}
    \centering
    \includegraphics[width=\hsize]{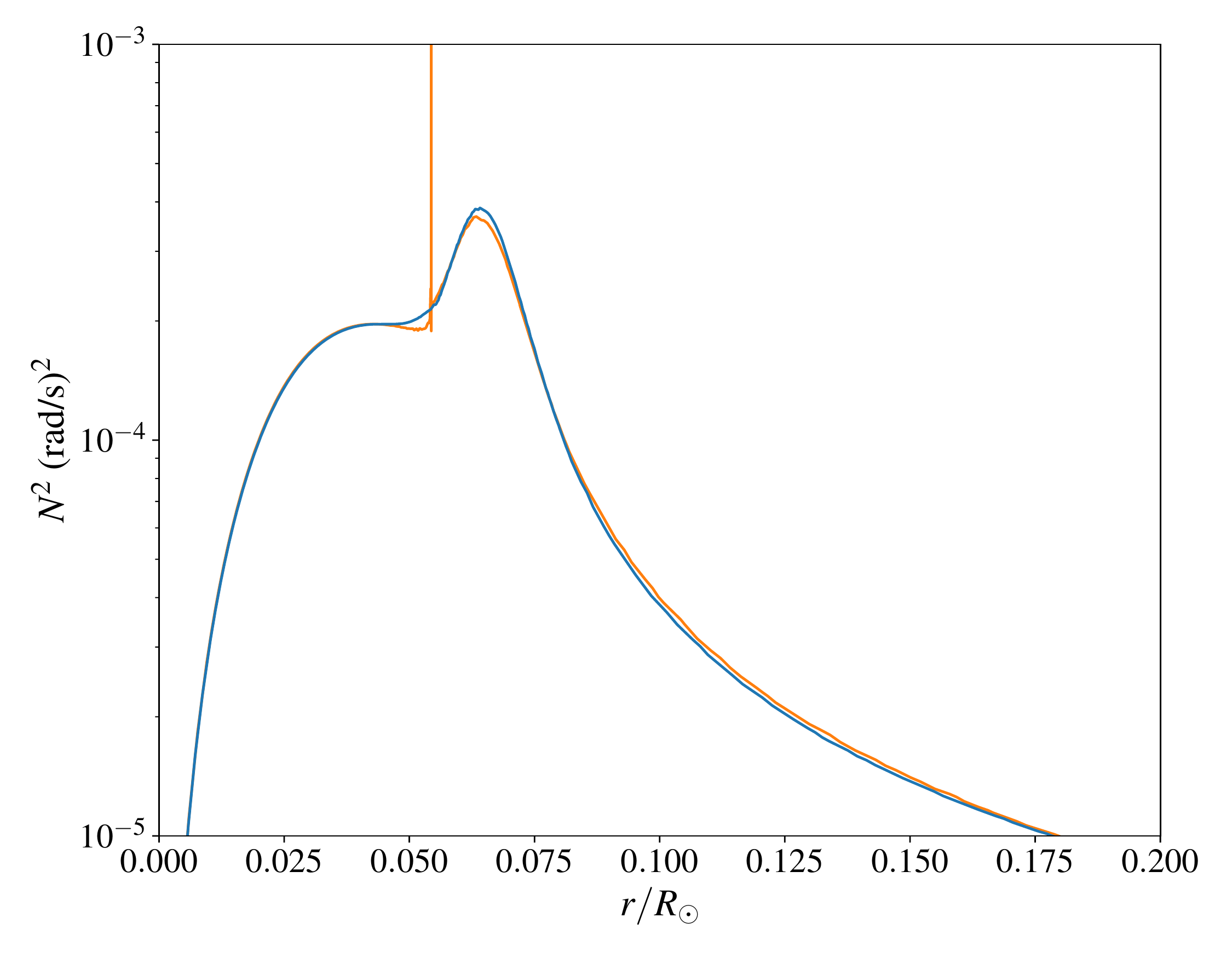}    \caption{\N profiles of the best models with gravitational settling and chemical diffusion (blue) and without those two processes (orange).}
    \label{profils_diffusion}
\end{figure}
The models presented in Sect.~\ref{results} of our study include both gravitational settling and chemical diffusion. Such processes, which happen to be necessary in order to correctly reproduce the helioseismic observations \citep{jcd93}, are expected to have an impact on our results for two main reasons. The first is the sinking during the main sequence of heavier elements because of the gravitational settling. This reduces the hydrogen abundance in the core and shortens the main sequence, which eventually decreases the age of models with same mean density. The second is the smoothing of the structure of the subgiant. High $\mu$-gradient peaks, like the one produced by the withdrawal of the convective core, are strongly smoothed out by the chemical diffusion, which impacts the mixed mode frequencies.
Thus, it is interesting to see how gravitational settling and chemical diffusion change the characteristics and the quality of the fit. We therefore modeled the star following a similar methodology, but without including those two processes. We found that the best model also has $\ov=0.15$, but provides a significantly worse agreement with the observations than the best diffusion model, with $\chi^2_{\mathrm{nodiff}} - \chi^2_{\mathrm{diff}} = 71$. It is more massive ($1.29\,\smass$) and older ($4.85\,$Gyr), as was expected from the fact that heavy elements do not sink during the MS. We note that the surface observables are less well reproduced, with a best model that is too cold ($5874\,$K, which represents $1.84$ times the observational uncertainty) and has too low a luminosity 
($5.0\,L_{\odot}$, which represents $4.3\,\sigma$). 

Moreover, similarly to what we found in Sect.~\ref{results}, the quality of the fit improves as $\ov$ increases for $\ov \leq 0.15$. However, this is less significant ($\chi^2_{\ov = 0} - \chi^2_{\ov = 0.15} = 24$). For higher values, the quality of the fit strongly worsens, in a comparable way to what has been found with gravitational settling and chemical diffusion.

Figure \ref{profils_diffusion} illustrates the differences in the \N profiles between the two best models, with and without both diffusive processes. One can see the high $\nabla_{\mu}$ peak (at $r=0.06\,R_{\odot}$) is, as expected, smoothed out by the chemical diffusion. Otherwise, the two profiles are remarkably similar, despite the different physics of the models, which highlights the robustness of the constraints coming from the mixed modes.

Finally, the effects of gravitational settling are expected to be somewhat counter-balanced in the envelope by radiative accelerations, which can have a significant impact on stars with masses greater than $1.1\,\smass$ \citep{deal18}. However, including the effects of radiative accelerations in the models increases the computational time of stellar evolution calculation by orders of magnitude, and it could not be afforded in the present study. To test the impact of this process on our modeling, we computed a model that takes into account radiative accelerations, with the same parameters as the best model for $\ov=0.15$. We obtained slightly different large separations ($\Delta \nu_{\rm rad} - \Delta \nu_{\rm norad} = 0.12 \, \mu\mathrm{Hz}$) but very similar frequencies, once normalized by the large separation. Radiative accelerations are therefore not expected to change the conclusions of this study.

\subsection{Helium degeneracy}
\label{Y0_degen}

\begin{figure}
    \centering
    \includegraphics[width=\hsize]{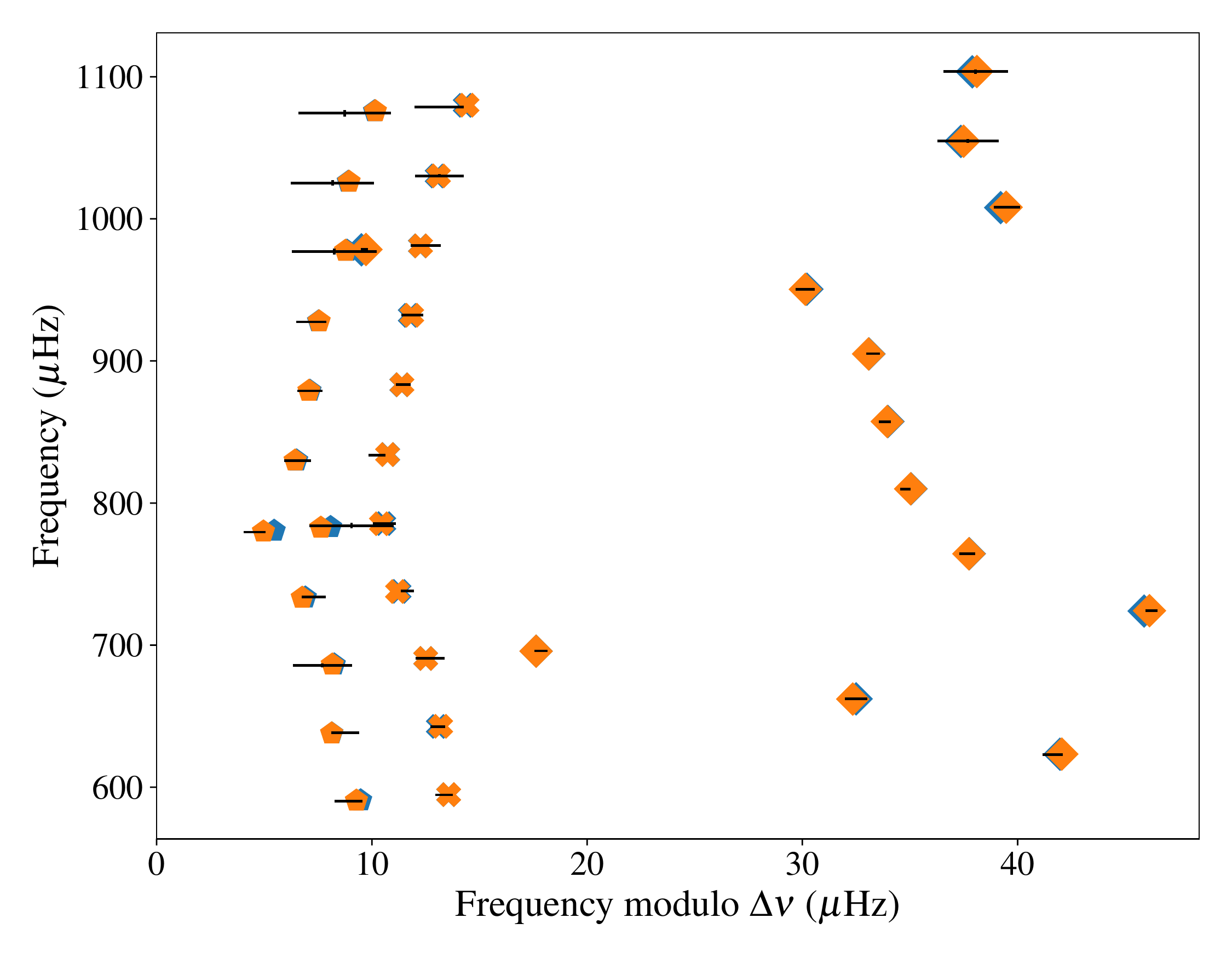}    \caption{Echelle diagrams of the best model (blue, $Y_0=0.28$) and the best model for $Y_0 = 0.24$ (orange). 3-$\sigma$ uncertainties from the observations are represented by black bars.}
    \label{echelle_degen}
\end{figure}
\begin{table}[]
    \centering
    \begin{tabular}{l c c}
        \hline \hline
        Param & $Y_0=0.24$ model & $Y_0=0.28$ model \\
        \hline
        $M$ ($\smass$) & 1.292 & 1.222 \\
        $R$ ($R_{\odot}$) & 2.152 & 2.109 \\
        Age (Gyr)         & 4.13 & 3.88 \\ 
        $L$ ($L_{\odot}$) & 5.95 & 5.89 \\
        $\teff$ (K)       & 6145 & 6192\\
        $\metal$ (dex)    & -0.098 & -0.072 \\
        $\conv$           & 1.70 & 1.73 \\
        $Y_0$            & 0.24 & 0.28 \\
        $\chi^2$         & 159 & 127 \\
        \hline
          
    \end{tabular}
    \caption{Characteristics of the two models of Fig.~\ref{echelle_degen}}
    \label{charac_degen}
\end{table}

To model KIC10273246, we initially performed optimizations with fixed $\ov$ and considering $Y_0$, $\conv$ and $\metal$ as free parameters. In this case, we observed an unwanted sensitivity of the $Y_0$ parameter to the guess value of our optimization process. This led us to the hybrid approach described in Sect.~\ref{modeling_process}, using optimizations with fixed values of $Y_0$ and varying $\metal$, $\conv$. We found that optimal models with different values of $Y_0$ indeed have surprisingly close frequencies, despite their wide range of mass. This is illustrated in Fig.~\ref{echelle_degen}, which shows the échelle diagrams of the best model with $Y_0=0.28$ (blue) and the best model with $Y_0=0.24$ (orange), both of which have $\ov=0.15$. Those models have quite different characteristics, as reported in Table~\ref{charac_degen}. However, their frequencies are almost indistinguishable, despite the very small uncertainties on the mode frequencies from \textit{Kepler} data. Only the g-dominated $l=2$ mode allows us to slightly favor the $Y_0=0.28$ model. Such degeneracy is related to the anti-correlation between mass and $Y_0$, that has been observed in MS stars (see e.g., \citealt{lebreton14}) as well as subgiant stars \citep{li20}. 
Additionally, we note that no monotonic behavior has been found between the age and $Y_0$.
We therefore conclude that the seismic modeling of subgiants, despite bringing strong constraints on the deep structure of the star, does not lift the degeneracy between $Y_0$ and the mass.

\subsection{Internal rotation} \label{sect_rotation}

We mentioned in Sect. \ref{sect_modefit} that a rotational splitting of $0.53 \pm 0.03\,\mu$Hz could be measured with a high level of significance for the $l=2$ mode at 779.4 $\mu$Hz. This is obviously not enough to probe the internal rotation in detail. However, since this mode is g-dominated, it can be used to place approximate constraints on the rotation in the core of the star.

Using our best-fit model from Sect. \ref{results}, we were able to compute the rotational kernel $K(r)$, which relates the splitting $\delta\nu_{\rm s}$ of this mode to the rotation profile $\Omega(r):$ 
\begin{equation}
    \delta\nu_{\rm s} = \int_0^R K(r)\Omega(r)/(2\pi) \,\hbox{d}r.
\end{equation}
This can be re-written as $\delta\nu_{\rm s} = K_{\rm g} \langle\Omega_{\rm g}\rangle + K_{\rm p} \langle\Omega_{\rm p}\rangle$, where $\langle\Omega_{\rm g}\rangle$ and $\langle\Omega_{\rm p}\rangle$ are average rotation rates in the g- and p-mode cavities, respectively, and $K_{\rm g}$ (resp. $K_{\rm p}$) corresponds to the integral of $K(r)$ in the g-mode (resp. p-mode) cavities. For the $l=2$ mode under study, using our best-fit stellar model we found that 84\% of the kernel energy is enclosed in the g-mode cavity, which confirms that the mode is indeed g-dominated. 

\cite{campante2011} found a clear rotational modulation in the \kepler\ light curve of KIC10273246. They thus estimated the surface rotation rate of the star to about 0.5 $\mu$Hz (rotation period of about 23 days). This value is comparable to the average rotational splitting of $0.45\pm0.02\,\mu$Hz that we obtained in this study. As mentioned in Sect. \ref{sect_modefit}, this average splitting is dominated by the contribution of p-dominated modes, to the extent that it essentially measures the envelope rotation rate. Taken together, these two measurements suggest a low rotation contrast within the p-mode cavity, which is in line with the conclusions of \cite{benomar15} and \cite{nielsen15} for main-sequence solar-like pulsators.

The splitting measured for the $l=2$ g-dominated mode is close to the rotation rate inferred for the envelope, which suggests a low amount of radial differential rotation in the star. If we take $\langle\Omega_{\rm p}\rangle/(2\pi) \approx 0.45\,\mu$Hz, we  obtain a core rotation rate of about 0.65 $\mu$Hz (rotation period of about 18 days). Clearly, more rotational splittings would be required to precisely measure the core rotation rate. However, our results indicate that KIC10273246 could be rotating nearly as a solid-body, like the two \kepler\ subgiants whose internal rotation profiles were recently measured (\citealt{deheuvels20}).

\section{Conclusion}
\label{conclusion}

In this study, we performed a seismic modeling of KIC10273246, a subgiant observed by Kepler, and obtained strong constraints on the size of its MS convective core. We chose this target because it exhibits two dipolar g-dominated modes, which we expected to bring stronger constraints on the internal structure. We extracted the mode parameters from the oscillation spectrum of the star using the full Kepler data set and thus updated the mode frequencies that were previously obtained by \cite{campante2011}. 

The seismic modeling of subgiants is notoriously complex. We here elaborated on the algorithm proposed by \cite{deheuvels11}. This method consists of a two-step approach. The purpose of the first step is to find the mass and age that match the large separation of p modes and the frequency of one g-dominated mixed mode. The second step optimizes the other free parameters ($\metal$, $Y_0$, $\conv$ and $\ov$) to reproduce the other frequencies as closely as possible. In this work, we improved this algorithm to make it more robust. This enabled us to perform a detailed seismic modeling of KIC10273246, with a particular emphasis on the investigation of the size of the MS convective core.

We found models in good agreement with the observations, with a reduced $\chi^2$ of 3.2 for the best model, and with surface observables that are reproduced to within less than 1 $\sigma$. One key result of this study is that models with core overshooting during the MS reproduce the observations significantly better, with an optimal value of $\ov = 0.15$. For higher values of $\ov$, the quality of the fit significantly worsens. We found that such models are very close to the TAMS. Their internal structure thus differs from that of the lower-$\ov$ solutions, and their seismic properties show strong mismatch with the observations. We tested the robustness of our conclusions by considering other choices for the input physics. No significant difference was found when modeling core overshooting as a diffusive process. Models computed without microscopic diffusion also favor models with $\ov=0.15$, albeit less significantly, and show a strong mismatch compared with the observations for higher values of $\ov$. However, they yield poorer agreement with the seismic and surface observables compared to the models computed with microscopic diffusion. This study thus confirms the high potential of young subgiants with mixed modes to measure the extent of the MS convective cores.

We also investigated the information conveyed by the mixed modes about the core structure. We showed that the combined knowledge of the large separation $\Delta\nu$ and the frequency of one g-dominated mixed mode is enough to estimate the central density $\rho_{\rm c}$ to a precision of about 1\%. This helps us understand why models with a greater amount of core overshooting ($\ov>0.15$) are not compatible with the observations. Because of their larger MS convective core, they have a higher $\rho_{\rm c}$ just after the end of the MS, and they thus reach the optimal central density closer to the TAMS. We then studied the roles of the different mixed mode frequencies in determining the profile of the Brunt-Väisälä frequency inside the star. While the first g-dominated mixed mode strongly constrains the thermal part, the second one helps constrain the part of the Brunt-Väisälä frequency that is dominated by the $\mu$-gradient. We therefore confirm that having access to two g-dominated mixed modes helps better characterize the Brunt-Väisälä profile.

Also, despite the strong constraints that were obtained on the internal structure, we noted the existence of a degeneracy between the stellar mass and the initial helium abundance $Y_0$. This degeneracy, which is already well known for MS stars (e.g., \citealt{lebreton14}), is not lifted by the mixed modes. We find that it is in fact the main source of uncertainties in the determination of the stellar parameters. This should be kept in mind when modeling subgiants. Current modeling techniques, such as traditional grid-based methods, tend to miss a significant fraction of the best-fit models because of the size of the mesh. In such conditions, the degeneracy between $Y_0$ and mass could be explored only partially, thus causing the uncertainties on the stellar parameters to be underestimated.

As a byproduct of this work, we obtained partial constraints on the internal rotation of KIC10273246. We were not able not measure individual rotational splittings for the dipolar mixed modes, but we obtained a splitting of $0.53\pm0.03\mu$Hz for the only g-dominated $l=2$ mixed mode in the spectrum of the star. Interestingly, this value is close to the surface rotation rate of $0.5\mu$Hz that was found for this star by \cite{campante2011} using photometric data from Kepler. This suggests that this star might be rotating nearly as a solid-body, similarly to the two young subgiants recently studied by \cite{deheuvels20}.

This work highlights the large potential of the seismic modeling of young subgiants to indirectly obtain constraints on the core structure of the star during the MS. The next step will be to use this method on a larger sample of stars drawn from the targets observed with Kepler and TESS, and therefore place quantitative constraints on the overshooting process in low-mass stars. The data from the upcoming PLATO mission \citep{plato} will add a large amount of potential targets for this type of analysis.

Moreover, we show in this study that detecting several g-dominated dipole modes places stronger constraints on the shape of the \N profile, and therefore on the $\mu$-gradient in the stellar core. It could thus be anticipated that more evolved subgiants, which show a larger number of g-dominated mixed modes, would be more favorable targets for our purpose. However, these stars are also further from the end of the MS, and a worry is that the chemical composition in the core might be less dependent on the properties of the MS core. We plan, therefore, to study this effect in a subsequent study.

\begin{acknowledgements}
We thank Morgan Deal for enlightening discussions about microscopic diffusion. We also thank the anonymous referee for comments that improved the clarity of this paper.
We acknowledge support from from the project BEAMING ANR-18-CE31- 0001 of the French National Research Agency (ANR) and from the Centre National d'Etudes Spatiales (CNES).
\end{acknowledgements}

\bibliographystyle{aa} 
\bibliography{biblio} 

\begin{appendix}
\section{Appendix A: Observed frequencies}
\begin{table}[h]
\begin{center}
\caption{Estimated mode parameters for KIC10273246.
\label{tab_seismic_param}}
\vspace{0.2cm}
\begin{tabular}{c c c c}
\hline \hline
\T\B $l$ & $\nu$ ($\mu$Hz) & $H$ (ppm$^2$/$\mu$Hz) & $\Gamma$ ($\mu$Hz)  \\
\hline
\T\B $0$ & $ 594.58\pm0.13$ & $10.9^{+ 2.8}_{- 2.2}$ & $ 1.4^{+ 0.4}_{- 0.3}$  \\
\T\B $0$ & $ 642.73\pm0.11$ & $10.5^{+ 3.3}_{- 2.5}$ & $ 1.2^{+ 0.5}_{- 0.3}$  \\
\T\B $0$ & $ 690.80\pm0.23$ & $ 7.9^{+ 1.7}_{- 1.4}$ & $ 3.5^{+ 1.0}_{- 0.8}$  \\
\T\B $0$ & $ 738.19\pm0.10$ & $18.4^{+ 2.7}_{- 2.3}$ & $ 1.7^{+ 0.2}_{- 0.2}$  \\
\T\B $0$ & $ 785.00\pm0.13$ & $17.3^{+ 2.1}_{- 1.8}$ & $ 2.2^{+ 0.2}_{- 0.2}$  \\
\T\B $0$ & $ 833.65\pm0.13$ & $18.9^{+ 2.4}_{- 2.1}$ & $ 2.4^{+ 0.3}_{- 0.3}$  \\
\T\B $0$ & $ 883.30\pm0.12$ & $18.7^{+ 2.3}_{- 2.1}$ & $ 2.2^{+ 0.2}_{- 0.2}$  \\
\T\B $0$ & $ 932.16\pm0.17$ & $13.0^{+ 1.6}_{- 1.5}$ & $ 2.8^{+ 0.3}_{- 0.3}$  \\
\T\B $0$ & $ 981.26\pm0.30$ & $ 8.9^{+ 1.4}_{- 1.2}$ & $ 4.9^{+ 1.0}_{- 0.8}$  \\
\T\B $0$ & $1030.30\pm0.38$ & $ 4.4^{+ 0.9}_{- 0.7}$ & $ 4.2^{+ 0.9}_{- 0.8}$  \\
\T\B $0$ & $1078.72\pm0.38$ & $ 3.2^{+ 1.0}_{- 0.8}$ & $ 3.0^{+ 1.2}_{- 0.9}$  \\
\T\B $1$ & $ 622.85\pm0.16$ & $ 6.2^{+ 1.8}_{- 1.4}$ & $ 1.6^{+ 0.5}_{- 0.4}$  \\
\T\B $1$ & $ 662.16\pm0.17$ & $ 6.3^{+ 1.3}_{- 1.0}$ & $ 2.4^{+ 0.5}_{- 0.4}$  \\
\T\B $1$ & $ 695.96\pm0.10$ & $11.1^{+ 5.7}_{- 3.8}$ & $ 0.7^{+ 0.5}_{- 0.3}$  \\
\T\B $1$ & $ 724.33\pm0.10$ & $12.9^{+ 2.6}_{- 2.1}$ & $ 1.5^{+ 0.3}_{- 0.3}$  \\
\T\B $1$ & $ 764.19\pm0.12$ & $13.0^{+ 1.7}_{- 1.5}$ & $ 2.6^{+ 0.3}_{- 0.3}$  \\
\T\B $1$ & $ 809.76\pm0.08$ & $23.7^{+ 3.4}_{- 3.0}$ & $ 1.7^{+ 0.2}_{- 0.2}$  \\
\T\B $1$ & $ 857.24\pm0.09$ & $19.6^{+ 2.6}_{- 2.3}$ & $ 2.0^{+ 0.2}_{- 0.2}$  \\
\T\B $1$ & $ 905.14\pm0.10$ & $16.3^{+ 2.1}_{- 1.8}$ & $ 2.3^{+ 0.3}_{- 0.2}$  \\
\T\B $1$ & $ 950.42\pm0.15$ & $ 9.1^{+ 1.2}_{- 1.1}$ & $ 3.0^{+ 0.4}_{- 0.3}$  \\
\T\B $1$ & $ 978.28\pm0.10$ & $11.9^{+ 5.2}_{- 3.6}$ & $ 0.5^{+ 0.3}_{- 0.2}$  \\
\T\B $1$ & $1008.23\pm0.21$ & $ 5.8^{+ 0.8}_{- 0.7}$ & $ 3.5^{+ 0.5}_{- 0.5}$  \\
\T\B $1$ & $1054.85\pm0.48$ & $ 2.4^{+ 0.4}_{- 0.4}$ & $ 6.2^{+ 1.3}_{- 1.1}$  \\
\T\B $1$ & $1103.65\pm0.50$ & $ 2.0^{+ 0.5}_{- 0.4}$ & $ 6.1^{+ 3.1}_{- 2.0}$  \\
\T\B $2$ & $ 590.15\pm0.21$ & $ 5.4^{+ 1.4}_{- 1.1}$ & $ 1.4^{+ 0.4}_{- 0.3}$  \\
\T\B $2$ & $ 638.42\pm0.22$ & $ 5.2^{+ 1.6}_{- 1.2}$ & $ 1.2^{+ 0.5}_{- 0.3}$  \\
\T\B $2$ & $ 685.80\pm0.46$ & $ 4.0^{+ 0.8}_{- 0.7}$ & $ 3.5^{+ 1.0}_{- 0.8}$  \\
\T\B $2$ & $ 733.83\pm0.19$ & $ 9.2^{+ 1.3}_{- 1.2}$ & $ 1.7^{+ 0.2}_{- 0.2}$  \\
\T\B $2$ & $ 779.53\pm0.03$ & $ 54.2^{+ 19.0}_{- 14.1}$ & $ 0.3^{+ 0.1}_{- 0.1}$  \\
\T\B $2$ & $ 784.03\pm0.65$ & $ 12.5^{+ 3.2}_{- 2.5}$ & $ 1.5^{+ 0.5}_{- 0.4}$  \\
\T\B $2$ & $ 829.97\pm0.21$ & $ 9.4^{+ 1.2}_{- 1.1}$ & $ 2.4^{+ 0.3}_{- 0.3}$  \\
\T\B $2$ & $ 878.97\pm0.20$ & $ 9.3^{+ 1.2}_{- 1.0}$ & $ 2.2^{+ 0.2}_{- 0.2}$  \\
\T\B $2$ & $ 927.47\pm0.23$ & $ 6.5^{+ 0.8}_{- 0.7}$ & $ 2.8^{+ 0.3}_{- 0.3}$  \\
\T\B $2$ & $ 976.97\pm0.66$ & $ 4.4^{+ 0.7}_{- 0.6}$ & $ 4.9^{+ 1.0}_{- 0.8}$  \\
\T\B $2$ & $1025.32\pm0.65$ & $ 2.2^{+ 0.4}_{- 0.4}$ & $ 4.2^{+ 0.9}_{- 0.8}$  \\
\T\B $2$ & $1074.32\pm0.72$ & $ 1.6^{+ 0.5}_{- 0.4}$ & $ 3.0^{+ 1.2}_{- 0.9}$  \\
\hline
\hline
\end{tabular}
\end{center}
\end{table}

\end{appendix}

\end{document}